%% file: main.tex
\documentclass[letterpaper,twocolumn,10pt]{article}
\usepackage{usenix-2020-09}

\usepackage[english]{babel}
\usepackage{tikz}
\usepackage[dvipsnames]{xcolor}
\usepackage{graphicx}
\usepackage{caption}
\usepackage{subfigure}
\usepackage[noabbrev,capitalise]{cleveref}
\usepackage[framemethod=TikZ]{mdframed}
\usepackage{xargs} 
\usepackage[colorinlistoftodos,prependcaption,textsize=tiny]{todonotes}
\usepackage{hyperref}
\usepackage{xspace}
\usepackage{multirow}
\usepackage{float}
\usepackage{nicefrac}
\usepackage{setspace}
\usepackage{url}
\usepackage{soul}
\usepackage{makecell}
\usepackage[shortcuts]{extdash}
\usepackage{wrapfig}


\newcommand{\systemname}{Peregrine\xspace}

\newif\ifimportant{}
\importanttrue{}

\makeatletter
	\if@todonotes@disabled
		
	\else
		
	\fi
\makeatother

\crefformat{section}{\S#2#1#3} 

\renewcommand{\arraystretch}{1.2}

\newcommand\takeaway[1]{
\begin{mdframed}[roundcorner=4pt,backgroundcolor=black!4]
	\textbf{Takeaway: } #1
\end{mdframed}
}

\newcommand\keyinsight[1]{
\begin{mdframed}[roundcorner=4pt,backgroundcolor=black!4]
	\textbf{Key Insight: } #1
\end{mdframed}
}

\newcommand*\emptycirc[1][1ex]{\tikz\draw (0,0) circle (#1);}
\newcommand*\halfcirc[1][1ex]{%
  \begin{tikzpicture}
  \draw[fill] (0,0)-- (90:#1) arc (90:270:#1) -- cycle ;
  \draw (0,0) circle (#1);
  \end{tikzpicture}}
\newcommand*\fullcirc[1][1ex]{\tikz\fill (0,0) circle (#1);} 

\begin{document}



\title{\large \bf Where to Sample: Rethinking ML-Based Malicious Traffic Detection at Terabit Scale}


\newcommand*{\ist}{\includegraphics[scale=0.038]{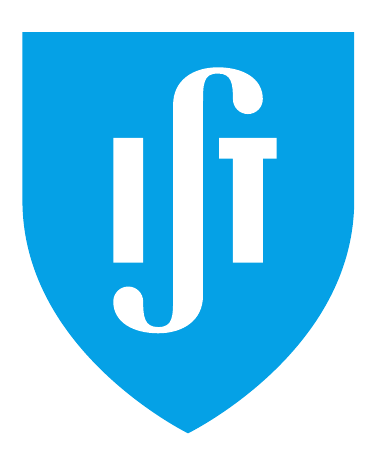}}
\newcommand*{\novafct}{\includegraphics[scale=0.005]{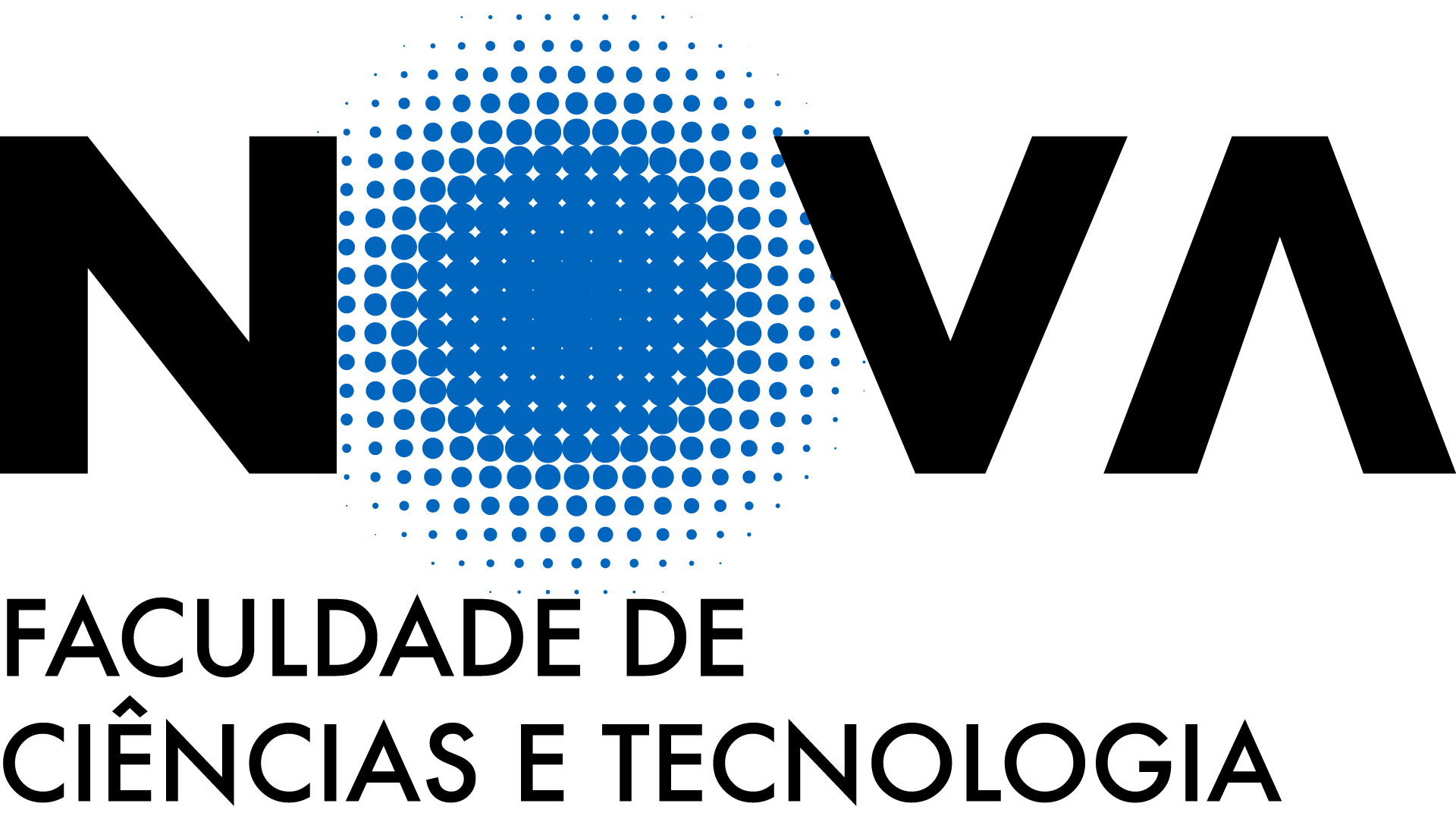}}
\newcommand*{\nyu}{\includegraphics[scale=0.004]{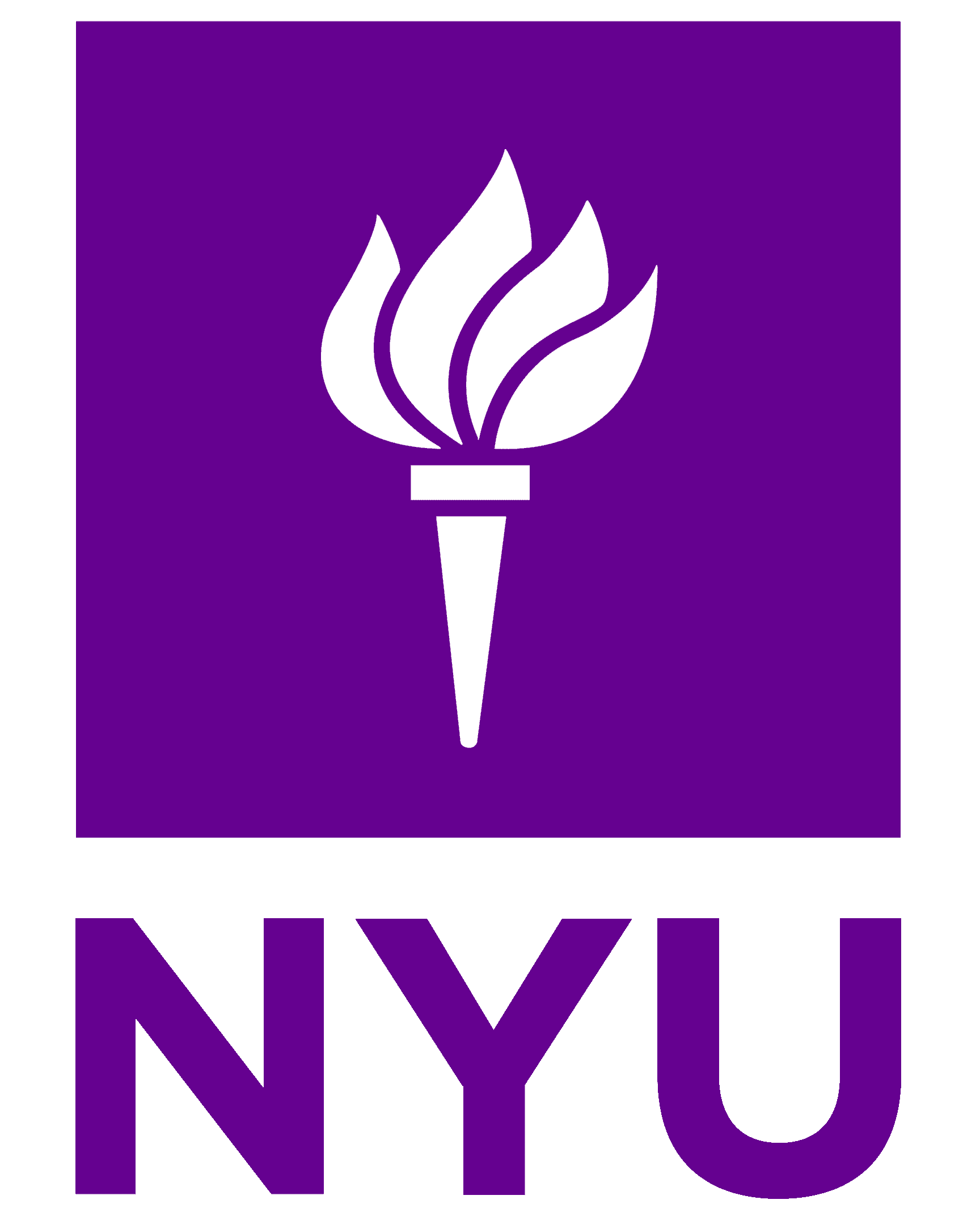}}

\author{
  {\rm João Romeiras Amado \ist \quad}
  {\rm Francisco Pereira\,\ist \quad}
	  {\rm David Pissarra\,\nyu \quad} \\
	  {\rm Salvatore Signorello\,\novafct \quad}
	  {\rm Miguel Correia\,\ist \quad}
	  {\rm Fernando M. V. Ramos\,\ist\quad} \\
	  {\small \ist\,INESC-ID, Instituto Superior Técnico, University of Lisbon \quad} \\
	  {\small \novafct\, NOVA University Lisbon\quad} \\
  {\small \nyu\,New York University\quad}
 }

\maketitle

\begin{abstract}
\input{./sections/abstract}
\end{abstract}

\input{./sections/introduction}
\input{./sections/motivation}
\input{./sections/threat_model}

\input{./sections/system_design}
\input{./sections/implementation}
\input{./sections/evaluation}
\input{./sections/discussion}
\input{./sections/conclusion_future_work}
\input{./sections/acknowledgements}

\bibliographystyle{plain}
\bibliography{references}

\newpage

\renewcommand{\thesubsection}{\Alph{subsection}}

\appendix

\input{./sections/appendix-implementation}
\input{./sections/appendix-eval-meth}
\input{./sections/appendix-kitsune}

\input{./sections/appendix-hv}

\input{./sections/appendix-planter}
\input{./sections/appendix-approx}
\input{./sections/appendix-train-sampl}
\input{./sections/appendix-runtime}
\input{./sections/appendix-resource-usage}
\input{./sections/appendix-efficiency}
\input{./sections/appendix-rw}

\end{document}

%% file: sections/abstract.tex

Machine Learning (ML)-based malicious traffic detectors can identify unseen attacks, including in encrypted traffic, but scaling them to Terabit rates remains challenging.
Packet sampling, commonly used to meet processing limits, can discard essential behavioral information and render detection ineffective.

We present \systemname, a hardware-software co-designed architecture built around a simple principle: \textit{compute features before sampling}.
\systemname updates ML features over all traffic at line rate in a programmable switch, then downsamples the feature records sent to software inference.
We instantiate this design with two detectors that impose distinct feature-computation requirements on the switch.
Across 84 attack traces from three datasets, we show that \textit{approximate features computed over all packets can outperform exact features computed from sampled packets}.
In our evaluation, \systemname preserves detection effectiveness while reducing feature-record rates by three orders of magnitude—even where pre-feature sampling causes detection to collapse.



%% file: sections/introduction.tex
\section{Introduction}
\label{sec:introduction}

An ideal Network Intrusion Detection System (NIDS) should simultaneously (\textbf{R1}) observe all network traffic at line rate, (\textbf{R2}) detect a wide range of attacks, including previously unseen ones, and (\textbf{R3}) maintain a consistently low false-positive rate.
Traditional rule- and signature-based NIDSs remain widely deployed because they offer excellent throughput---e.g., Pigasus achieves up to 100 Gbps on a single server~\cite{pigasus}---and low false positives, but this performance relies on predefined signatures and payload-oriented inspection.
As a result, such systems struggle in environments where traffic is predominantly encrypted---the norm in today's networks~\cite{cisco-whitepaper,barradas2021flowlens}---and may miss attacks that do not match existing signatures; moreover, the reported throughputs remain below the Terabit-scale regime we target.

To address these limitations, a growing body of work proposes anomaly-based Machine Learning (ML) NIDSs that model normal traffic behavior and identify attacks as deviations from learned patterns~\cite{mirsky2018kitsune,whisperFu2021,buczak2016survey,chaabouni2019network,fu2023hypervision}.
By operating on traffic statistics rather than payload signatures, these systems can analyze encrypted traffic and detect previously unseen attacks.
Recent advances in ML have demonstrated impressive detection accuracy and low false-positive rates in controlled and moderately scaled settings~\cite{mirsky2018kitsune,whisperFu2021,fu2023hypervision}.

\noindent
\textbf{Problem.} Despite their promise, ML-based NIDSs face a fundamental scalability barrier when deployed in modern high-speed networks.
The combined throughput of feature computation and ML inference remains far below the traffic rates of modern high-speed networks.
As a result, established online detectors report throughputs ranging from Mbps to tens of Gbps~\cite{mirsky2018kitsune,whisperFu2021,fu2023hypervision}, far below the Terabit-per-second data rates increasingly common in backbone and data center networks~\cite{singh2015jupiter,poutievski2022jupiter}.
\textit{The bottleneck is not model accuracy, but end-to-end system throughput.}

\begin{figure}[t]
\centering
    \begin{minipage}{.49\textwidth}
        \centering
        \subfigure[Kitsune]{\includegraphics[width=.49\linewidth]{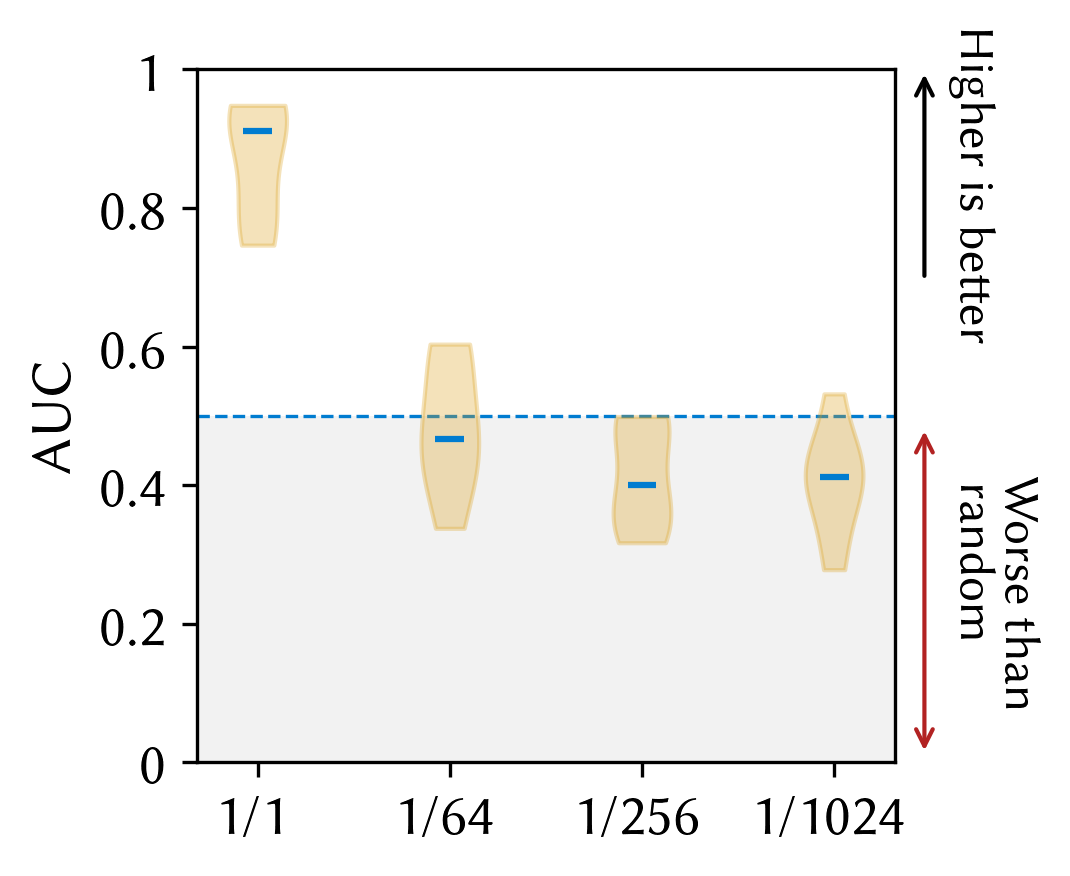}}
        \subfigure[HyperVision]{\includegraphics[width=.49\linewidth]{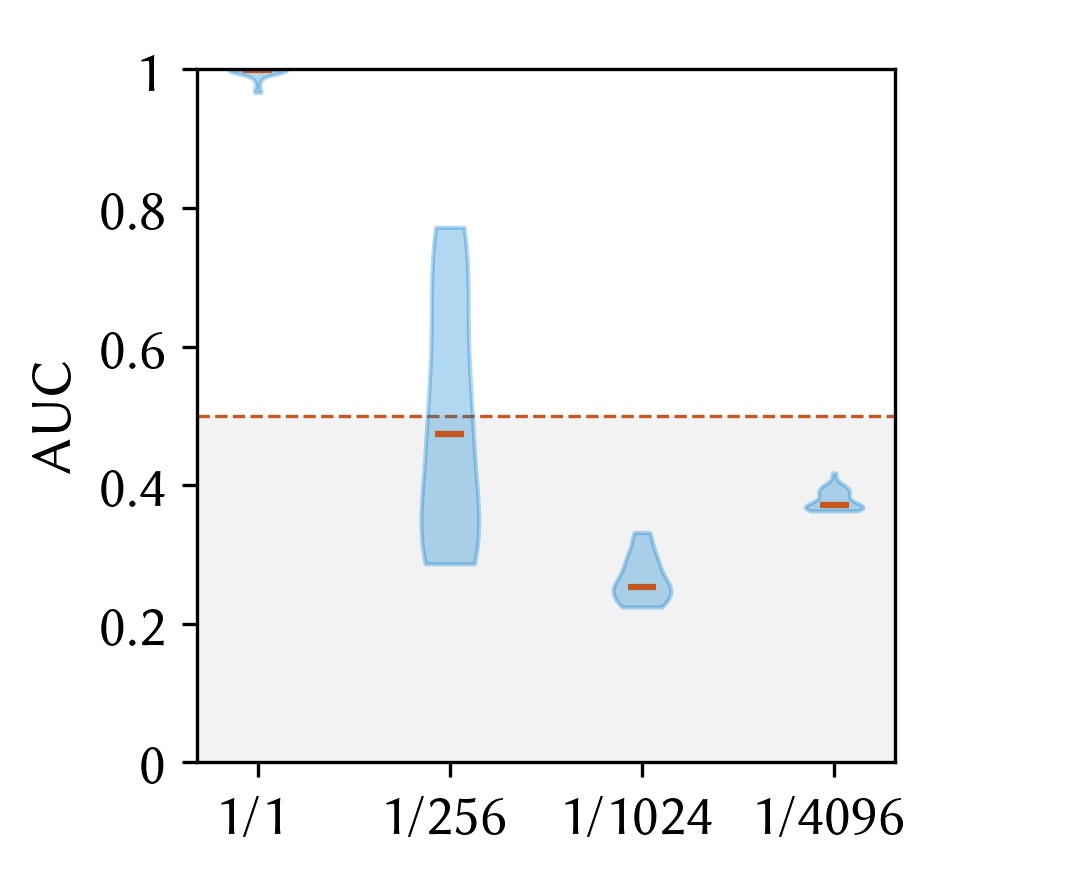}}
        \captionsetup{width=.9\linewidth}
        \caption{Detection performance (AUC) of Kitsune and HyperVision~\cite{mirsky2018kitsune,fu2023hypervision} on a combined set of 84 attack traces from three datasets~\cite{mirsky2018kitsune,sharafaldin2018toward,fu2023hypervision}, across sampling rates. Aggressive pre-feature packet sampling substantially degrades detection performance.}
        \label{fig:problem}
    \end{minipage}
\end{figure}

To bridge this gap, a common approach is to sample traffic before feature computation.
This design choice fundamentally limits visibility, undermining the very signals ML-based detectors rely on.
As Figure~\ref{fig:problem} shows, the detection performance of two ML-based detectors~\cite{mirsky2018kitsune,fu2023hypervision} collapses under aggressive pre-feature sampling rates, often performing worse than random guessing\footnote{AUC measures how well a detector separates malicious from benign traffic across all decision thresholds, making it a threshold-independent measure of detection effectiveness. AUC = 0.5 corresponds to chance-level ranking.}.
The degradation observed in these two distinct detection pipelines highlights \textit{the loss of behavioral information caused by sampling raw packets prior to feature-extraction}.
Aggressive pre-feature sampling thus compromises (\textbf{R1}), and can undermine (\textbf{R2}) even when the underlying model performs well on unsampled traffic.\\

\keyinsight{The central scalability question for ML-based malicious traffic detection is not only how much to sample, but also \textit{where sampling occurs} in the detection pipeline.}

Prior work has largely focused on optimizing packet processing, feature-extraction, or inference, pushing software- and NIC-based systems to increasingly higher throughputs~\cite{Ghasemirahni2024Fajita,Gerry2022Retina,whisperFu2021,fu2023hypervision}.
However, these systems still leave a substantial gap to Terabit-scale traffic rates: closing this gap requires either additional processing resources or some form of traffic reduction.
We argue that the key remaining challenge is \textit{architectural}: determining where traffic reduction should occur in the detection pipeline.

Prior work provides important building blocks for addressing this architectural challenge: Sonata~\cite{gupta2018sonata} partitions telemetry queries between switches and software, while FastFE and SuperFE~\cite{bai2020fastfe,zhang2025superfe} accelerate ML feature-extraction across programmable devices.
In this paper we build on these foundations by treating sampling placement as a design decision whose impact on detection effectiveness must be evaluated alongside throughput.

\begin{figure}[t]
    \centering
    \includegraphics[width=.99\linewidth]{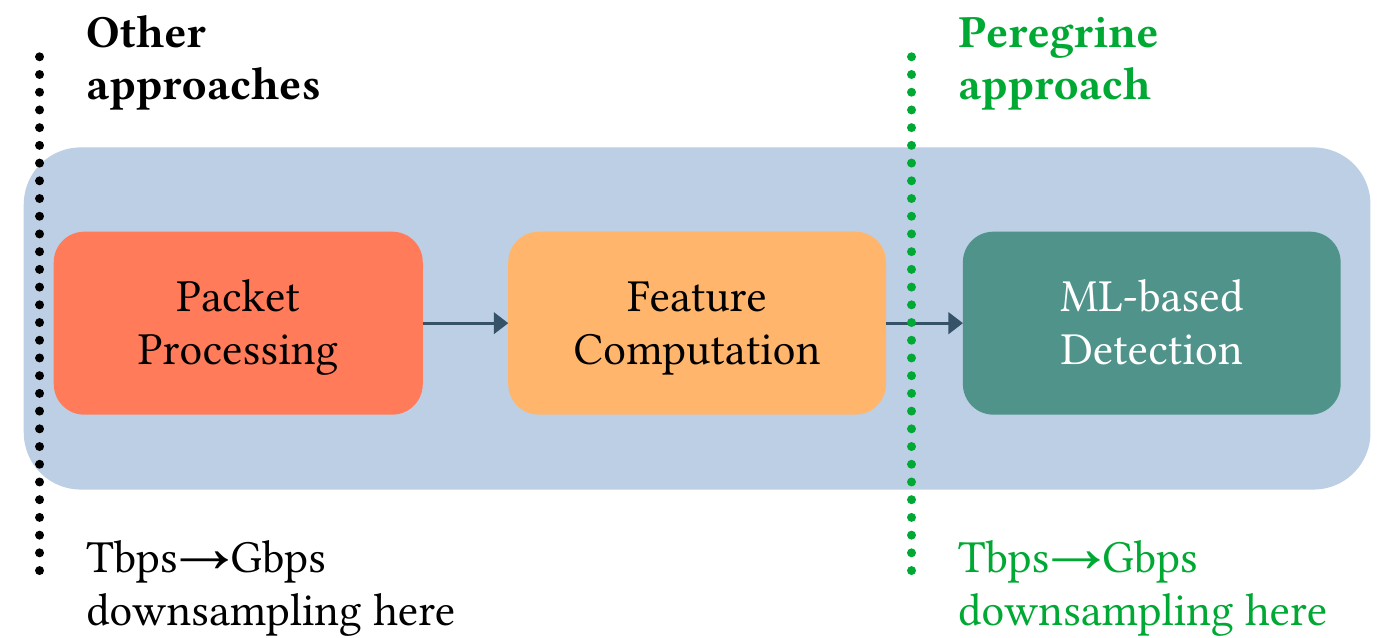}
    \captionsetup{width=.99\linewidth}
    \captionof{figure}{Sampling placement in ML-based malicious traffic detection. Pre-feature packet sampling reduces downstream processing load by discarding packets before feature-extraction. \systemname~instead computes ML features over all traffic at line rate and applies downsampling only after feature computation, preserving packet contributions to feature state before reducing the inference workload.}
    \label{fig:downscaling}
\end{figure}

\noindent
\textbf{Rethinking sampling placement.}
Our design follows a simple principle: \textit{compute ML features over all traffic at line rate, and apply downsampling only after feature computation}.
Figure~\ref{fig:downscaling} illustrates this distinction, contrasting conventional pre-feature sampling with our proposed sampling-after-feature-computation design.

To realize this principle, we introduce \systemname, a hard\-ware-software (HW-SW) co-designed architecture that leverages programmable switch data planes to compute rich ML features at Terabit speeds.
The switch acts as a \textit{visibility amplifier}, maintaining behavioral feature state before reducing the stream of records sent to the ML detector.
Our target is a high-bandwidth monitoring environment with bounded numbers of concurrently tracked flows, where traffic volume, rather than flow cardinality, is the dominant scaling challenge.
Examples motivating this operating regime include AI training fabrics~\cite{Qian2024AlibabaHPN} and monitoring at rack granularity~\cite{Benson2010Wild}.
By offloading feature computation from software servers into the data plane, \systemname enables a substantially more efficient realization of ML-based detection.

\noindent
\textbf{~\systemname~Overview.}
\systemname splits the detection pipeline across two platforms with complementary strengths.
The programmable switch data plane~\cite{bosshart2013forwarding,intelTofinos} performs per-packet feature computation at line rate, while a commodity server executes full-fledged ML inference on compact feature records.
At the boundary between these two domains, we apply post-feature downsampling: reducing the rate of feature records passed to software inference while continuing to update feature state from every packet.
We instantiate this architecture in two concrete systems, \systemname-K and \systemname-H, using Kitsune~\cite{mirsky2018kitsune} and HyperVision~\cite{fu2023hypervision} as representative detectors whose distinct feature requirements pose different implementation challenges in the switch data plane.
These two realizations demonstrate our sampling principle across different feature-extraction pipelines, rather than tying it to a particular ML model.

This HW-SW co-design yields three key advantages.
First, the preservation of traffic visibility during feature computation.
Features are computed per-packet, over all traffic, at line rate, preserving behavioral information that pre-feature sampling would discard.
Second, leveraging powerful ML models.
Inference remains in software, avoiding the need to simplify ML models to fit switch memory and computational constraints, which can compromise detection effectiveness~\cite{xiong2019switches,busse2019pforest,zhou2023efficient,yan2024brain,zheng2024planter,zhang2025quark}.
Third, cost and energy efficiency.
Offloading feature computation to the switch
reduces the server resources needed to process traffic at line rate, creating opportunities to lower both operational cost and power consumption.
~\systemname uses switch resources to preserve traffic visibility during feature computation, with a configurable feature-record emission rate to match the workload to the processing capacity of the software backend.

\begin{table*}[]
	\footnotesize
	\begin{singlespace}
    \centering
    \begin{tabular}{|l|c|c|c|c|c|c|}
		\hline
		\multicolumn{1}{|c|}{\textbf{Class}} & \textbf{\makecell[c]{Systems}} & \textbf{\makecell[c]{Execution Model}} & \textbf{\makecell[c]{Traffic\\ Visibility}} & \textbf{\makecell[c]{Feat.\\Expressiv.}} & \textbf{\makecell[c]{Detection\\Scope}} & \textbf{\makecell[c]{Decision Granularity}} \\ \hline
		\makecell[l]{Rule-based\\middleboxes} & \cite{roesch1999snort},\cite{paxson1999bro},\cite{pigasus} & \makecell[c]{Middlebox rules / DPI} & \textcolor{orange}\halfcirc & \textcolor{ForestGreen}\fullcirc & \textcolor{orange}\halfcirc & \makecell[c]{Per-packet / Per-flow} \\ \hline
		\makecell[l]{ML-based\\middleboxes} & \makecell[c]{\cite{invariantBartos2016},\cite{whisperFu2021},\cite{wang2022enidrift},\\ \cite{mirsky2018kitsune},\cite{fu2023hypervision}} & Off-path SW ML & \textcolor{orange}\halfcirc & \textcolor{ForestGreen}\fullcirc & \textcolor{ForestGreen}\fullcirc & \makecell[c]{Per-flow / Windowed} \\ \hline
		\makecell[l]{Attack-specific\\ data plane} & \makecell[c]{\cite{liu2021jaqen},\cite{zhang2020poseidon}\\\cite{alcoz2022aggregate},\cite{misa2024leveraging}} & \makecell[c]{Specialized\\ in-network logic} & \textcolor{ForestGreen}\fullcirc & \textcolor{orange}\halfcirc & \textcolor{orange}\halfcirc & Per-packet \\ \hline
		In-switch ML & \makecell[c]{\cite{zhou2023efficient},\cite{yan2024brain}\\\cite{zheng2024planter}\cite{zhang2025quark}} & Data plane ML & \textcolor{ForestGreen}\fullcirc & \textcolor{red}\emptycirc & \textcolor{ForestGreen}\fullcirc & Per-packet \\ \hline
		\makecell[l]{\textbf{\systemname}\\\textbf{(HW-SW co-design)}} & \textbf{\systemname} & \textbf{\makecell[c]{Data plane feature-extraction\\+ Off-path SW ML}} & \textcolor{ForestGreen}\fullcirc & \textcolor{ForestGreen}\fullcirc & \textcolor{ForestGreen}\fullcirc & \textbf{\makecell[c]{Per-epoch\\(sampled feature records)}} \\ \hline
    \end{tabular}
	\end{singlespace}
	\caption{While prior systems trade off feature expressiveness, detection scope, or traffic visibility at Terabit scales, \systemname~combines expressive feature-extraction with full line-rate visibility by shifting sampling to occur \textit{after} feature computation.\ (\textcolor{ForestGreen}\fullcirc~= Full \textcolor{orange}\halfcirc~= Partial  \textcolor{red}\emptycirc~= Narrow)}
	\label{tab:nids-comparison}
\end{table*}

In summary, this paper makes the following contributions:\\
\textbf{Sampling placement.} We show that preserving traffic visibility can matter more for detection than feature precision: approximate features computed over all packets can outperform exact features computed from sampled packets.\\
\textbf{Architecture.} We introduce \systemname, a HW--SW co-de\-signed architecture that computes ML features in the data plane before reducing the records sent to software inference.\\
\textbf{Hardware realization.} We realize this approach in two concrete systems integrating distinct detectors, using hardware-aware approximations and state-management mechanisms to fit their feature computations into programmable switches.\\
\textbf{Evaluation.} We evaluate \systemname across 84 attack traces from three datasets, showing that detection effectiveness can be preserved under downsampling by three orders of magnitude---even when pre-feature sampling causes detection to collapse.

\textit{This work does not raise ethical issues.}

%% file: sections/motivation.tex
\section{Background and motivation}
\label{sec:motivation}

In this section, we review the state-of-the-art in malicious traffic detection and the architectural limitations that prevent existing systems from scaling to Tbps networks.
We then motivate in-network feature computation and summarize the key challenges in realizing \systemname.

\subsection{Malicious traffic detection}
\label{subsec:RW}

Signature- or rule-based NIDSs~\cite{roesch1999snort,paxson1999bro,pigasus} remain widely deployed in operational networks due to their maturity and predictable performance.
However, encryption limits payload inspection, while attacks that do not match existing signatures can evade detection.

Anomaly-based ML detectors model normal traffic behavior and identify attacks as deviations from learned patterns, enabling detection of both encrypted and previously unseen attacks~\cite{buczak2016survey,chaabouni2019network,mirsky2018kitsune,whisperFu2021}.
However, these benefits come at the cost of significantly higher computational complexity.

Recent works implement attack-specific detection logic in the data plane~\cite{liu2021jaqen,zhang2020poseidon,alcoz2022aggregate,misa2024leveraging}, demonstrating Tbps-scale throughput but targeting narrow attack classes such as DDoS.

Other systems execute ML models directly in programmable switches~\cite{xiong2019switches,busse2019pforest,zhou2023efficient,yan2024brain,zheng2024planter,zhang2025quark}, but the available memory, arithmetic operations, and pipeline stages constrain the models and features they can support.

Table~\ref{tab:nids-comparison} summarizes this landscape.
The detector classes in the table face different constraints: limited feature expressiveness (in-switch ML), restricted detection scope (attack-specific data plane systems), or reduced visibility when pre-feature sampling is used to sustain processing limits (software and middlebox-based ML).
\systemname~combines line-rate feature computation with software inference, computing features over all traffic at line rate and applying downsampling only \emph{after} feature computation.
Complementary feature-extraction frameworks, including FastFE~\cite{bai2020fastfe} and SuperFE~\cite{zhang2025superfe}, support downstream detectors.
FastFE completes feature computation on a server, while SuperFE computes features on SmartNICs from metadata batched by the switch.
In contrast, \systemname performs feature computation in the switch before downsampling the records sent to software inference.
These and other related works are discussed in Appendix~\ref{appendix:rw}.

\subsection{Motivation and opportunity}
\label{subsec:motivation}

Our motivation to develop \systemname is threefold.
First, ML-based detectors have demonstrated strong effectiveness against encrypted and previously unseen attacks while maintaining low false-positive rates~\cite{ mirsky2018kitsune,whisperFu2021,fu2023hypervision}.

The second motivation is scalability.
Reported throughputs include 150 Mbps for Kitsune, 15 Gbps for Whisper, and approximately 80 Gbps for HyperVision, although these results use different workloads and experimental setups~\cite{mirsky2018kitsune,whisperFu2021,fu2023hypervision}.
These rates remain well below Terabit-scale traffic rates.
Bridging that gap requires additional processing resources or traffic reduction. However, applying packet sampling before feature computation can substantially degrade detection effectiveness, as~\cref{fig:problem} illustrates.
We therefore argue that the key challenge is architectural: determining where traffic reduction should occur in the detection pipeline.

Finally, programmable switches sustain Tbps throughput while supporting stateful in-network computation~\cite{intelTofinos}, motivating a HW-SW co-design that moves feature computation into the data plane.

\subsection{Target Switch Architecture}
\label{subsec:pisa}
The targets of our design are programmable switching platforms capable of sustaining Terabit-scale traffic.
We focus on the Protocol Independent Switch Architecture (PISA) \cite{PISA}, embodied in Intel's Tofino switching ASICs~\cite{intelTofinos}.
Our implementations target Tofino and Tofino 2, with the latter supporting an aggregate switching capacity up to 12.8 Tbps, but the architectural principles should also apply to other high-end programmable switching ASICs (e.g., Broadcom Trident-4~\cite{TridentSwitch}).

A PISA-based switch consists of a programmable parser and deparser, ingress and egress pipelines composed of match\-/action stages.
Each pipeline stage contains limited compute resources---simple ALUs and on-chip memory---and executes at a fixed clock rate to guarantee deterministic per-packet processing latency.
Stateful processing is supported through register arrays, enabling limited cross-packet state.

These architectural constraints impose strict limits on per-packet computation, restricting the models and feature computations that can execute entirely in the switch pipeline.
The available arithmetic operations are hardware-dependent and substantially more restricted than on a general-purpose processor, and each stateful memory element can typically be accessed at most once per packet.

\systemname~is explicitly designed around these constraints.
It leverages the strengths of PISA switches---deterministic line-rate processing and efficient stateful operations---to offload packet processing and feature computation to the data plane, while leaving ML inference to software running on commodity servers.

\subsection{Design Challenges}
\label{subsec:challenges}

Designing a malicious traffic detection system that operates effectively at Terabit scale raises several system-level challenges rooted in the architectural constraints of programmable switch data planes (\cref{subsec:pisa}).
While switches offer unparalleled throughput and visibility, limited per-packet computation, constrained state access, and restricted memory capacity fundamentally shape how traffic detection can be realized.
As a result, any viable design must balance efficiency, functionality placement, and detection fidelity under extreme throughput.

\noindent
\textbf{Efficiency at scale.}
ML-based detectors rely on computationally intensive feature-extraction and inference.
Processing every packet in software makes feature computation a recurring CPU and I/O cost as traffic rates increase, potentially becoming a bottleneck in the overall system.
Offloading this work can reduce server requirements, but consumes the more limited switch resources, (see~\cref{subsec:system-implications} and~\cref{appendix:efficiency}).
The challenge is to implement useful features within the switch's arithmetic, stage, and memory budgets while retaining their value for detection.


\noindent
\textbf{Division of functionality.}
A central challenge is how to partition the detection pipeline across heterogeneous platforms.
Switches excel at simple, massively parallel per-packet processing, while servers support complex analytics and ML inference.
Placing all logic in the data plane limits model expressiveness, while placing it entirely in software requires scaling processing resources with the monitored traffic.
An effective design must therefore partition functionality according to platform strengths, explicitly accounting for switch constraints that restrict inference complexity but support streaming feature computation.

\noindent
\textbf{Sampling and aggregation.}
Computing features before sampling preserves information from packets that would otherwise be discarded.
However, the resulting feature records must still fit the software backend's processing capacity.
As such, it is necessary to control record emission while retaining useful temporal and behavioral information, accounting for the effect of reduced emissions on detection coverage and delay.



\noindent
\textbf{State management.}
Packet-processing capacity and flow-state capacity impose distinct scaling constraints.
Our target are high-bandwidth environments with bounded flow populations, motivated by AI training fabrics and rack-level monitoring~\cite{Qian2024AlibabaHPN,Benson2010Wild}. 
More diverse traffic mixes, such as the CAIDA traces examined in our evaluation (Section~\ref{subsec:system-implications}) highlight the importance of sizing state for the monitored workload.
Maintaining behavioral features across packets requires an efficient use of the limited switch memory.
The challenge is to orchestrate feature state collection and coordinate updates and expiration within the switch's restricted memory-access budget, sustaining line-rate processing for the targeted flows.

\noindent
\textbf{Implications.}
These challenges motivate an architecture that combines line-rate feature computation in the data plane with ML inference in software.

%% file: sections/threat_model.tex




%% file: sections/system_design.tex

\section{Architecture and Design}
\label{sec:design}

We now present the design of the \systemname \textit{approach}, which operationalizes our key insight on sampling placement via a hardware-software co-designed architecture.
We instantiate this approach in two concrete systems that integrate fundamentally different ML-based detectors.

We begin in~\cref{subsec:rationale} by describing how sampling placement reconciles line-rate feature computation with limited software inference capacity.
We then present the architecture (\cref{subsec:overview}) and the threat model (\cref{sec:threatmodel}), followed by two instantiations---\systemname-K and \systemname-H---that apply the same principles to distinct detection pipelines.
For each system, we describe data plane feature computation (\cref{subsec:peregrine-data-plane}) and control plane ML detection (\cref{subsec:peregrine-control-plane}).

\subsection{Design rationale and guiding principles}
\label{subsec:rationale}

Starting from the standard ML-based NIDS pipeline (Figure~\ref{fig:downscaling}), we reconsider where each stage should execute to reconcile Terabit-scale traffic rates with substantially slower ML processing.

\textit{Packet Processing (PP).} Packet parsing and forwarding remain in the switch data plane.

\textit{Feature Computation (FC).} ML-based detectors extract flow- and packet-level features (e.g., header fields and flow aggregates)~\cite{mirsky2018kitsune,whisperFu2021}.
Crucially, many ML features can be maintained incrementally through streaming per-packet updates, making them a natural fit for programmable data planes.

\textit{ML-based Detection (MD).} Executing ML inference in the switch requires models compatible with its available arithmetic, memory, and pipeline resources~\cite{xiong2019switches,busse2019pforest,zhou2023efficient,yan2024brain}.
Executing ML inference in software on commodity servers allows the use of complex models without imposing the computational and memory constraints of the switch data plane on their inference algorithms.

This division yields two guiding principles.

\textit{Principle \#1: Compute ML features per packet in the switch data plane.}
A PISA pipeline can maintain the counters and aggregates that define ML features at line rate, ensuring that every packet contributes to the feature space.

\textit{Principle \#2: Downsample only feature records sent to the ML detector.}
When the incoming feature record rate exceeds the software backend's processing capacity, record emission must be reduced or additional backend resources provisioned.
\systemname~therefore emits compact feature records at configurable granularity, downsampling only record emission rather than feature computation.
Aggregation maintains behavioral information across packets; post-feature downsampling controls how frequently that information is passed to inference.
Updating feature state for every packet does not necessarily imply that every packet or flow receives a separate detection decision.

\subsection{Architectural Overview}
\label{subsec:overview}

\systemname's detection pipeline spans a programmable switch and a commodity server.
We instantiate \systemname with Kitsune~\cite{mirsky2018kitsune} and HyperVision~\cite{fu2023hypervision}, referred to as \systemname-K and \systemname-H, to demonstrate the design across distinct feature-extraction requirements.

\systemname~decomposes detection across the switch and server.
The switch performs packet parsing and per-packet feature computation at line rate, incrementally updating \textit{feature atoms}---stateful counters and lightweight aggregates---from which higher-level ML features are derived.
Based on well-defined conditions (e.g., epochs or flow termination), it emits compact feature records to a commodity server, where full-fledged ML inference is performed.
This separation leverages complementary platform strengths: deterministic line-rate execution in the switch and expressive ML inference in software.

Crucially, \systemname applies traffic reduction \emph{after} feature computation.
Every packet contributes to the feature state, so the ML detector operates on summaries derived from the full traffic stream.
This preserves packet contributions before traffic reduction, in contrast to computing features from a pre-sampled packet stream.

\subsection{Threat Model}
\label{sec:threatmodel}


We assume that the network infrastructure---including the programmable switch and the middlebox server---is trusted and uncompromised.
The middlebox server runs ML-based detectors that either (i) are trained on benign traffic or (ii) operate without an explicit training phase.
We study how feature computation and sampling placement affect detection of malicious traffic, including attacks absent from the detector's training data.
Attackers may generate malicious traffic from compromised hosts.
Our evaluation does not establish robustness against training-data poisoning, adaptive evasion, or deliberate exhaustion of monitoring state.
The supported operating regime is bounded by the resident feature-state capacity and software processing budget described in~\cref{subsec:system-implications}.
\systemname targets monitoring and detection through exported feature records, rather than guaranteed prevention of an attack's first damaging packet.

\subsection{Feature Computation in the Data Plane}
\label{subsec:peregrine-data-plane}

This section describes how \systemname realizes data plane feature computation in practice, instantiating the principles introduced in~\cref{subsec:rationale}.
We present two concrete systems---\systemname-K and \systemname-H---that integrate fundamentally different ML-based detectors but share the same architectural pattern: per-packet feature updates at line rate in the data plane, followed by controlled emission of compact feature records to software.

\subsubsection{\systemname-K: Statistical Feature Computation}
\label{subsec:peregrine-k-data-plane}

\begin{figure*}[t]
    \centering
    \hspace{-1em}
    \subfigure[\systemname-K] {\includegraphics[width=.49\linewidth]{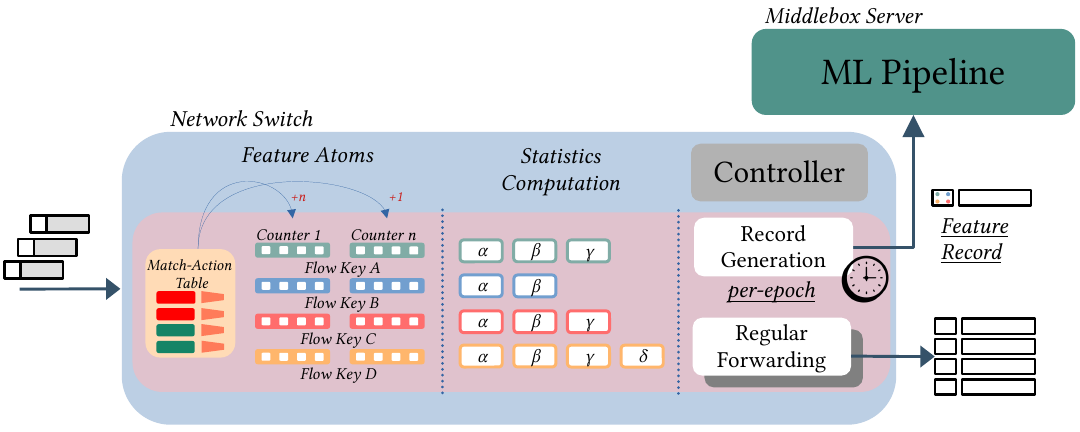}}
    \hspace{1em}
    \subfigure[\systemname-H]{\includegraphics[width=.49\linewidth]{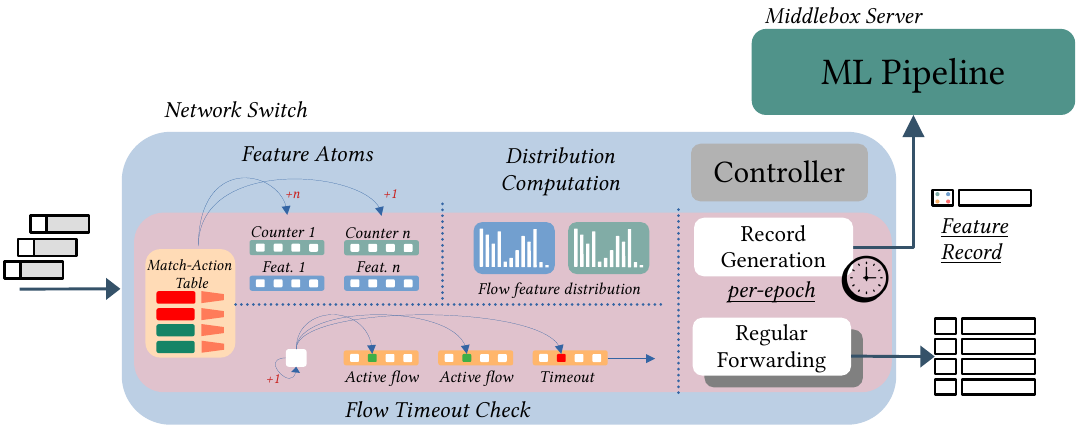}}
    \caption{\systemname: Feature computation in the data plane. Main operations per-packet and per-epoch.}
    \label{fig:peregrine-arch-data}
\end{figure*}

Figure~\ref{fig:peregrine-arch-data}(a) shows \systemname-K's packet processing pipeline.
Anon\-Arch-K demonstrates how statistical features originally computed in software~\cite{mirsky2018kitsune} can be incrementally maintained in the data plane without sacrificing line-rate performance.

Each incoming packet updates a small set of stateful counters (the ``feature atoms'') in the data plane at line rate.
Feature computation is structured as a streaming pipeline, decomposing higher-level features into simple PISA-compliant per-stage operations.

\noindent
\textbf{Feature Atoms.}
To compute flow-level features incrementally (e.g., mean or variance of packet sizes), \systemname-K maintains a small set of per-flow counters in the data plane---\textit{packet count (w)}, \textit{linear sum of packet sizes (LS)}, and \textit{squared sum of packet sizes (SS)}.
These atoms are sufficient to derive a wide range of higher-level features required by the ML backend, while remaining compatible with switch constraints.

\noindent
\textbf{Temporal Decay.}
To emphasize recent traffic behavior, Anon\-Arch-K applies an exponential decay to feature atoms before updating them.
This decay mechanism enables the detector to capture temporal dynamics---an essential property for identifying many classes of attacks---without storing per-packet histories.
Because switches cannot perform arbitrary exponentiation, \systemname-K implements decay using approximate techniques that trade numerical precision for line-rate execution.
Limited state-access resources prevent updating all decay instances on every packet.
\systemname-K therefore updates one decay instance per packet, rotating among instances across successive packets (more details in \cref{sec:implementation} and \cref{appendix:implementation}).

\noindent
\textbf{Statistics computation.}
Using the maintained feature atoms, \systemname-K computes both unidirectional and bidirectional statistics (\cref{appendix:implementation}), including measures such as mean, standard deviation, and correlation.
Bidirectional statistics are restricted to flow keys that naturally represent communication channels (e.g., 5-tuples), reflecting practical resource constraints.
Bidirectional statistics introduce dependencies between state associated with opposite traffic directions.
\systemname-K uses duplicated feature atoms to accommodate these dependencies within the pipeline (~\cref{appendix:implementation}).
Some arithmetic operations are approximated to fit the switch pipeline; these approximations aim to preserve qualitative feature behavior that is sufficient for downstream ML detection.
We evaluate the effect of these approximations on detection effectiveness in ~\cref{appendix:approx}.

\noindent
\textbf{Configuration and Sampling Semantics.}
\systemname-K's data plane is configured at compile time to support multiple flow keys (e.g., MAC source, IP source, 5-tuple), allowing operators to trade generality for resource usage.
The epoch length determines the rate at which feature records are generated and therefore directly controls the effective post-feature sampling rate.
Shorter epochs produce more feature records and less aggressive downsampling, while longer epochs increase traffic reduction by emitting fewer records. 
Importantly, this rate defines post-feature sampling, not packet sampling.
Regardless of the sampling rate, all packets continue to update feature state.
A sampling rate of 1:1024 means one feature record is emitted per 1024 packets, as determined by a global packet counter.
Each feature record is carried in a packet whose payload contains the approximate traffic statistics listed in Table~\ref{tab:primitives} of~\ref{sec:impl-peregrine-kitnet}.
At each emission event, the record contains the statistics associated with the triggering packet's aggregation keys.

\subsubsection{\systemname-H: Feature Distributions and Flow Dynamics}
\label{subsec:peregrine-h-data-plane}

\systemname-H instantiates the same architectural principles using a fundamentally different feature model inspired by HyperVision~\cite{fu2023hypervision}.
Figure~\ref{fig:peregrine-arch-data}(b) illustrates its packet processing pipeline.
In addition to scalar statistics, \systemname-H maintains compact per-flow feature distributions capturing packet-size and timing behavior.

\noindent
\textbf{Feature atoms.}
For each packet, \systemname-H updates a small set of per-flow feature atoms, including packet count, byte count, TCP flag counts, and inter-arrival time statistics.
In addition, it maintains fixed-size histograms (bins) for selected features such as packet length and inter-arrival time.
These bins are updated per packet, allowing the switch to approximate feature distributions incrementally.
This approach captures richer behavioral signals than scalar statistics, while remaining compatible with the data plane execution model.

\noindent
\textbf{Flow Timeout and Record Generation.}
\systemname-H emits feature records containing the maintained per-flow statistics and distributions when flows terminate or epochs expire.
Incoming packets advance timeout checks over the flow-state table.
Partitioning state across register banks allows updates and timeout checks to proceed without violating state-access constraints.
As with \systemname-K, only the emission of feature records is subject to downsampling; all packets are incorporated into the feature state.
We present additional details in \cref{sec:implementation} and \cref{appendix:implementation}.

\subsection{ML-based Detection}
\label{subsec:peregrine-control-plane}

\systemname~deliberately keeps ML-based detection in software, running on a commodity server, in order to preserve model expressiveness and adaptability.
This placement avoids adapting the selected inference algorithms to switch resource constraints, as required by in-switch ML solutions.
~\cref{subsec:eval-detection-planter} compares the resulting detection performance with the evaluated in-switch ML configurations.


The software backend consumes exported feature records, maintains any additional state required by the detector, and produces anomaly scores or classifications.
Its integration therefore depends on the supplied features and their temporal and sampling semantics.

Upon receiving a feature record, the server updates any required local state and feeds the record to the ML inference pipeline.
The configured record-emission rate controls the workload presented to the backend, while feature-state updates continue at line rate.

\subsubsection{\systemname-K}
\label{subsec:peregrine-k-control-plane}

\systemname-K integrates the KitNET anomaly detection framework~\cite{mirsky2018kitsune} as its ML backend.
KitNET is an unsupervised neural network composed of a hierarchy of autoencoders trained to reconstruct feature vectors representing benign traffic behavior.
Each received feature record is mapped to a fixed-dimensional input vector that is partitioned across multiple autoencoders.
The reconstruction error---measured using the root mean squared error (RMSE) between the input vector and the reconstructed output---is used as anomaly score.
Feature records that deviate significantly from the learned baseline result in high reconstruction error and are flagged as potentially malicious.

\systemname does not modify the detector itself; instead, by shifting sampling to occur after feature computation, it enables KitNET to operate on feature summaries derived from all traffic, with the goal of preserving the statistical signals required by this unsupervised ML model even under limited inference throughput.

\subsubsection{\systemname-H}
\label{subsec:peregrine-h-control-plane}

\systemname-H integrates HyperVision's ML-based detection pipeline~\cite{fu2023hypervision}, which models traffic behavior through graph-based representations of flow interactions.
Upon receiving feature records, the server constructs an interaction graph whose vertices represent network addresses and whose edges represent flows.
HyperVision applies a two-stage clustering process---combining density-based clustering (DBSCAN) and centroid-based clustering (K-Means)---to separate benign from anomalous traffic components.

This instantiation highlights that \systemname~is not limited to scalar or statistical feature vectors, but can also support richer relational and structural ML models.
The same architectural separation applies: HyperVision processes exported feature records in software, while post-feature downsampling controls the volume of records presented to the backend.

\takeaway{
Together, \systemname-K and \systemname-H demonstrate a shared architectural pattern across distinct feature-extraction pipelines: \textit{maintain feature state at line rate in the switch, then control record emission to software inference}.
Their implementations expose different arithmetic, state-access, and record-generation challenges, which we address in~\cref{sec:implementation}.}

%% file: sections/implementation.tex

\begin{table}[]
\centering
\footnotesize
\setlength{\tabcolsep}{3pt}
\renewcommand{\arraystretch}{1.08}
\begin{tabular}{|l|l|l|}
\hline
\multicolumn{1}{|c|}{\textbf{Requirement}} & \multicolumn{1}{c|}{\textbf{Constraint}} & \multicolumn{1}{c|}{\textbf{Mechanism}} \\ \hline
\textbf{K:} Derived statistics & Limited arithmetic & \begin{tabular}[c]{@{}l@{}}Bit-shift approximations,\\ hardware math units\end{tabular} \\ \hline
\textbf{K:} Multiple decay scales & Reg. access budget & \begin{tabular}[c]{@{}l@{}}Cycle updates across\\ decay instances\end{tabular} \\ \hline
\textbf{K:} Bidirectional statistics & \begin{tabular}[c]{@{}l@{}}Single access per-reg.\\ for each packet\end{tabular} & \begin{tabular}[c]{@{}l@{}}Duplicate atoms to\\ access both directions\end{tabular} \\ \hline
\textbf{H:} Flow update/expiration & \begin{tabular}[c]{@{}l@{}}Updates/timeout checks \\ compete for reg. access\end{tabular} & \begin{tabular}[c]{@{}l@{}}Update a reg. bank,\\ exp. check the others\end{tabular} \\ \hline
\end{tabular}
\caption{Mapping feature requirements to switch-compatible
mechanisms. K and H denote \systemname-K and \systemname-H.
Implementation details appear in Appendix~\ref{appendix:implementation}.}
\label{tab:implementation-mechanisms}
\end{table}

\section{Implementation Overview}
\label{sec:implementation}

We implemented \systemname on a real programmable switch and a commodity server, validating that sampling-after-feature-computation is feasible under the constraints of today's hardware.
Our prototype realizes a single architectural approach through two concrete systems, \systemname-K and \systemname-H, which implement distinct feature-extraction pipelines within the same hardware-software architecture.
Both systems share the same hardware-software partitioning: a P4 data plane program running on an Intel Tofino switch performs per-packet feature computation, while a software backend running on a commodity server receives feature records and executes ML-based detection.

The \systemname prototype consists of a few thousand lines of P4 for packet parsing, per-packet feature updates, epoch handling, and feature-record generation, together with server-side code for record ingestion and ML-based detection.
\systemname-K reuses the original KitNET pipeline implemented in Python, while \systemname-H integrates a C++ implementation derived from HyperVision's graph-based detection pipeline.
Both retain the detectors' inference algorithms while replacing their feature-extraction frontends.

The data plane implementation targets the Tofino Native Architecture (TNA), a realization of the PISA model used by Intel's Tofino switching ASICs.
We support both Tofino 1 (TNA) and Tofino 2 (T2NA), which differ primarily in available pipeline depth, state capacity, and total throughput.
\systemname-K includes implementations for both Tofino 1 and Tofino 2, while \systemname-H targets Tofino 2.
On Tofino 2, feature computation fits within a single pipeline.
On Tofino 1, \systemname-K updates feature atoms in the first pipeline and sends only packets selected for feature-record generation through an additional pipeline to complete the derived statistics.

Implementing \systemname raises several non-trivial challenges.
First, the switch pipeline does not natively support several arithmetic operations required by the feature computations, including division and square root, thus requiring hardware-compatible approximations.
Second, strict pipeline constraints---including limited state and a single access to each register array per packet---challenge the realization of temporal decay, bidirectional statistics, and flow-level aggregation.
Third, high-speed record generation and flow timeout detection must be performed without disrupting line-rate packet processing, despite the different feature semantics of \systemname-K and \systemname-H.

We address these challenges through hardware-aware feature computation and state management.
In \systemname-K, bit-shift arithmetic and hardware math units approximate operations unsupported by the pipeline's standard ALUs.
A cyclic update strategy handles multiple decay factors, while duplicated feature atoms across stages enable bidirectional statistics under register-access constraints.
In \systemname-H, partitioning flow state across independent register banks allows each packet to update its own flow while advancing timeout checks in other banks.
These mechanisms preserve per-packet updates to feature state while reducing the rate of records exported to software.

Table~\ref{tab:implementation-mechanisms} summarizes how we map the two feature-extraction pipelines to switch hardware.
These mechanisms combine arithmetic approximations with state-access scheduling, allowing packets to update feature state before record emission is reduced.
We evaluate their effect on detection performance in ~\cref{sec:evaluation} and Appendix~\cref{appendix:approx}.
We make the P4 data plane implementations of \systemname publicly available~\cite{anonarchrepo}.

Together, these techniques enable line-rate feature computation in the switch for two distinct detection pipelines, with record emission configured to match software processing capacity.
~\cref{appendix:implementation} details the implementations and their hardware-specific trade-offs.

%% file: sections/evaluation.tex
\section{Evaluation}
\label{sec:evaluation}

This section evaluates whether \systemname's hardware-software co-design enables effective ML-based malicious traffic detection at Terabit scale.
Our evaluation asks whether computing features before sampling can sustain effective malicious traffic detection under orders-of-magnitude traffic reduction, where pre-feature packet sampling fails, and what resources are needed to realize this design at Terabit scale.

We first evaluate detection effectiveness under varying sampling regimes, comparing \systemname-K and \systemname-H against their respective software-based baselines, Kitsune~\cite{mirsky2018kitsune} and HyperVision~\cite{fu2023hypervision}, as well as against representative in-switch ML approaches (\cref{subsec:eval-detection-kitsune}-\cref{subsec:eval-detection-planter}).
We then assess the practical implications of \systemname in terms of scalability, switch resource usage, communication overhead, and cost and energy efficiency, in \cref{subsec:system-implications}, with additional details and analyses provided in the appendices.

\begin{figure*}[t]
\centering
	\begin{minipage}[t]{.33\textwidth}
		\centering
		\includegraphics[width=.95\linewidth]{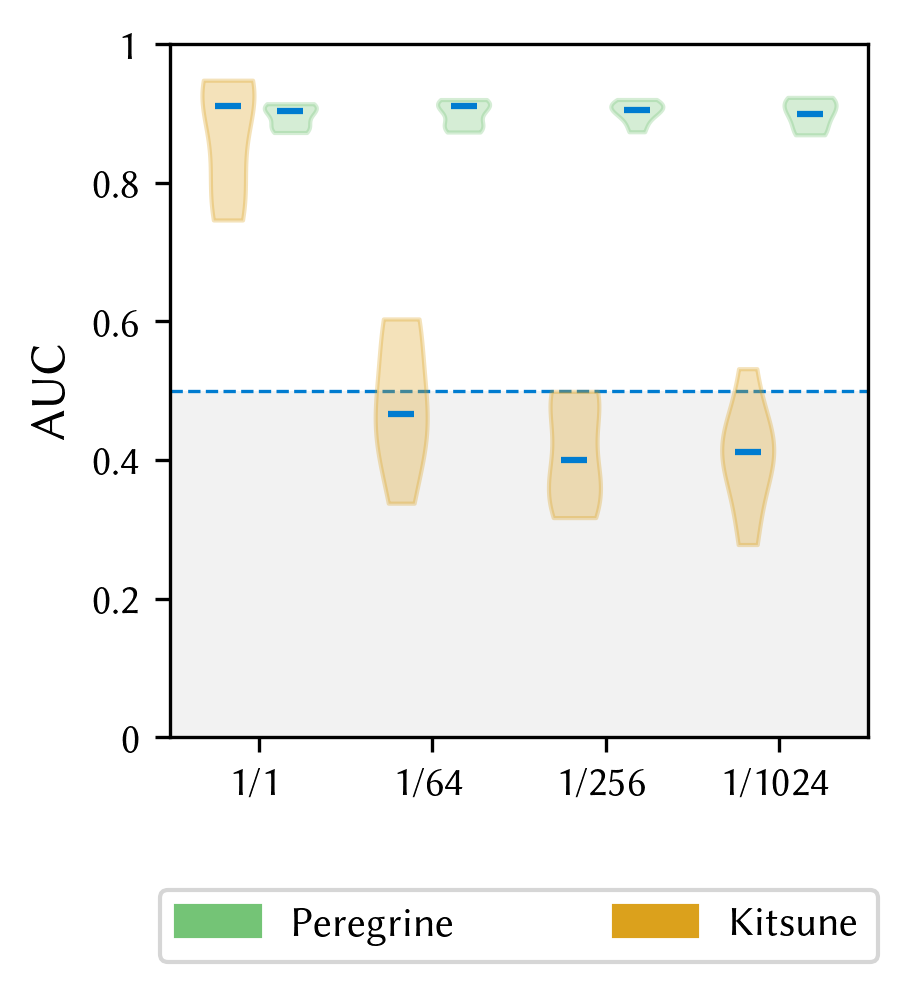}
		\captionsetup{width=.9\linewidth}
		\captionof{figure}{AUC across sampling rates (on x-axis): Kitsune and \systemname-K.}
		\label{fig:auc-kitnet-violin}
	\end{minipage}%
	\begin{minipage}[t]{.33\textwidth}
		\centering
		\includegraphics[width=.95\linewidth]{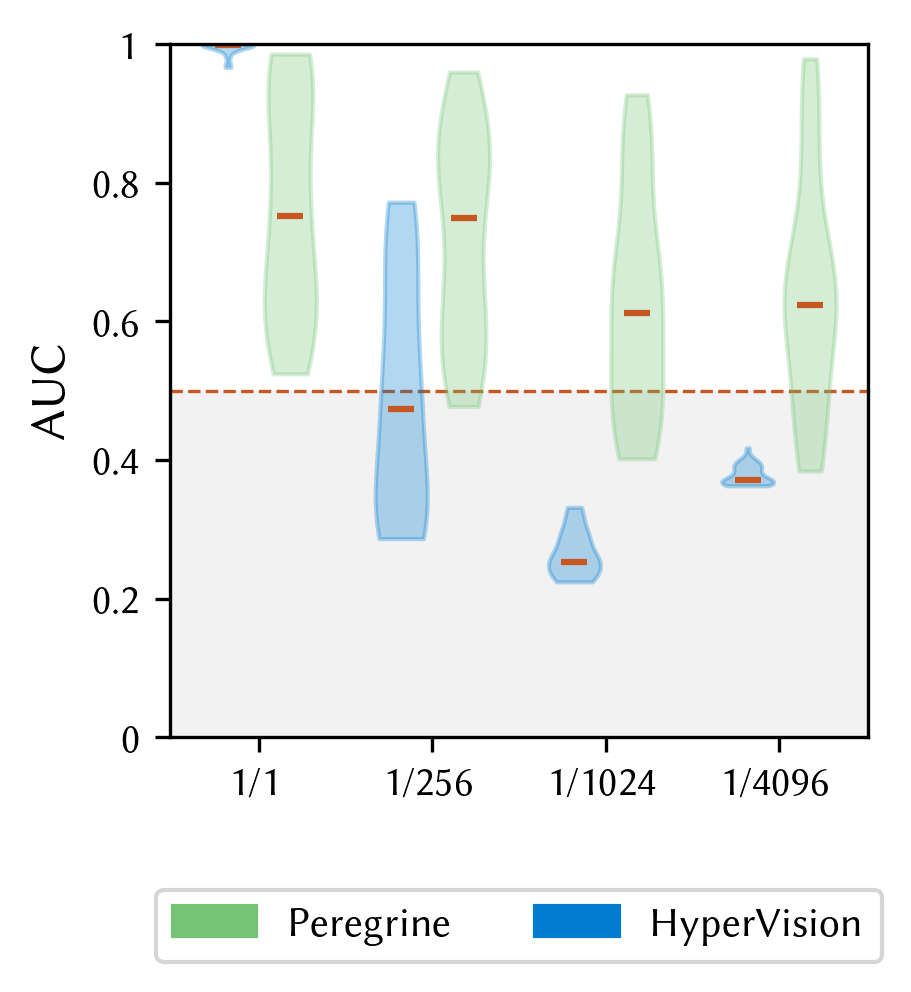}
		\captionsetup{width=.9\linewidth}
		\captionof{figure}{AUC across sampling rates (on x-axis): HyperVision and \systemname-H.}
		\label{fig:auc-hv-violin}
	\end{minipage}
	\begin{minipage}[t]{.33\textwidth}
		\centering
		\includegraphics[width=.95\linewidth]{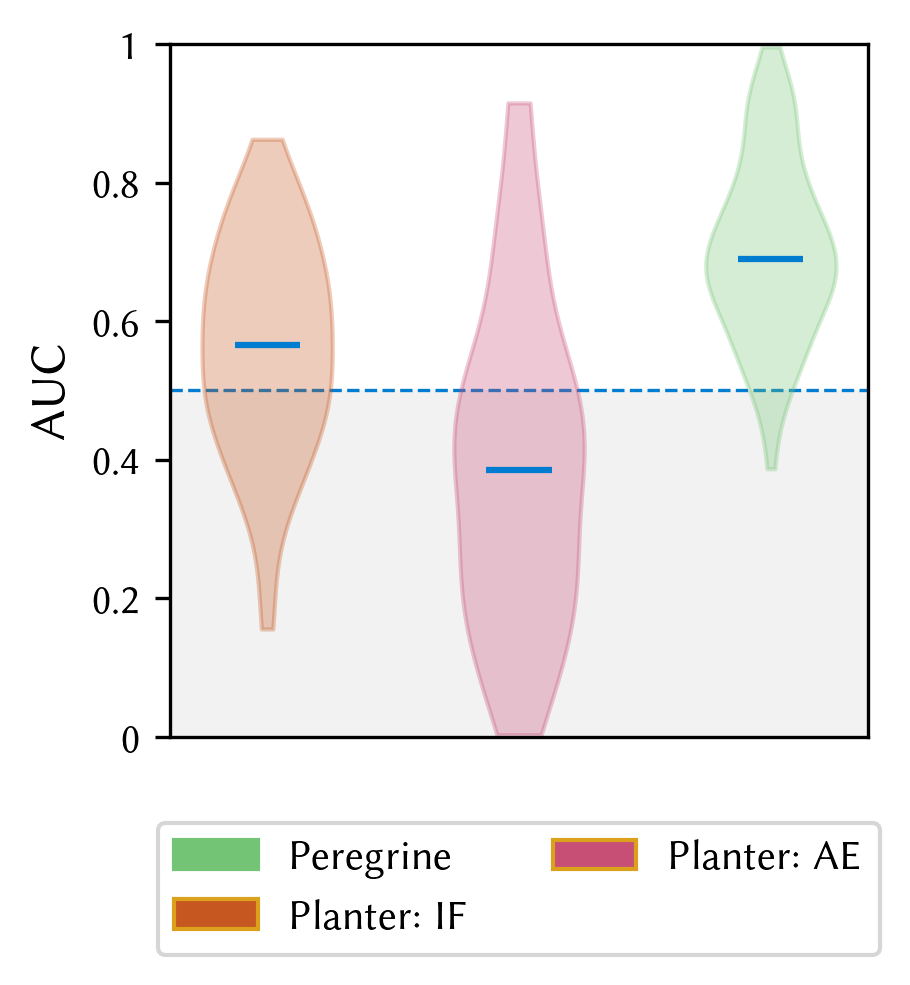}
		\captionsetup{width=.9\linewidth}
		\captionof{figure}{AUC for Planter unsupervised models and \systemname-K.}
		\label{fig:auc-planter-violin}
	\end{minipage}
\end{figure*}

\subsection{Setup and Methodology}

Our evaluation covers 84 attack traces drawn from 3 publicly available datasets widely used in prior work.
Specifically, we use the Kitsune dataset~\cite{mirsky2018kitsune}, which contains 9 attacks, the HyperVision dataset~\cite{fu2023hypervision}, with 69 attacks, and 6 attacks from CIC-IDS-2018~\cite{cic-datasets}.
For \systemname-K and Kitsune, we follow the original Kitsune methodology~\cite{mirsky2018kitsune} and train the detector using benign traffic preceding the attack period.
For \systemname-H and HyperVision, we follow the methodology described in~\cite{fu2023hypervision}.
Attack traffic is excluded from training.
The evaluation therefore tests detection without attack-specific training examples.
Unless otherwise stated, training uses unsampled benign traffic, and sampling is introduced during evaluation.
Appendix \cref{appendix:train-sampl} examines the sensitivity of \systemname-K to sampling during training.

Our testbed comprises a programmable switch and two commodity servers connected through 100 Gbps links.
Each server has dual-socket Intel Xeon Gold 6226R CPUs, 96 GB of DRAM, and Intel E810 NICs.
One server generates traffic, while the other receives feature records and executes ML inference.
We replay attack traces at their original rates when measuring detection performance, preserving their temporal characteristics.
The runtime performance is evaluated separately using a DPDK-based packet generator.
\cref{appendix:eval-meth} provides further details.

To evaluate scalability, we vary the traffic reduction required to
bridge Terabit-scale packet processing and software-based ML inference.
For the tested software baselines, this reduction is achieved via pre-feature
packet sampling.
For \systemname-K, a rate of 1:$k$ exports one feature
record every $k$ packets, using a global packet counter, while
every packet continues to update feature state.
In \systemname-H, the feature record is sent according to two conditions: sampling rate and flow timeout.
Every $k$ packets, based on the active sampling rate, all flows flagged as timed out during their respective timeout detection check are added to a feature record that is sent to the server backend.
Recall that the flow timeout detection is a process performed for each pipeline execution on all register banks except the one matching the current packet, as described in~\cref{appendix:implementation}.
Varying these rates lets us examine how detection
effectiveness changes as fewer packets or feature records reach
the software backend.
We relate the required reduction to
backend processing capacity in~\cref{subsec:system-implications}.

To maximize fairness, we use the same datasets originally employed in the evaluation of Kitsune and HyperVision.
Consequently, the maximum evaluated sampling rate is constrained by the available trace length: Kitsune is evaluated up to 1:1024, while the longer HyperVision traces allow evaluation up to 1:4096.
Detection effectiveness is reported using the Area Under the ROC Curve (AUC), a threshold-independent metric that captures the trade-off between true-positive and false-positive rates under varying sampling conditions.
We compute AUC over the observations scored by each detector: RMSE (Kitsune, \systemname-K) and loss function score (HyperVision, \systemname-H).
For Kitsune and \systemname-K, each scored packet and exported feature record, respectively, contributes one observation to the AUC calculation.
For HyperVision and \systemname-H, each scored flow contributes one observation.
Ground-truth labels are assigned using the dataset's flow labels.
Observations receiving no score are excluded; AUC therefore measures discrimination among scored observations rather than detection coverage over all traffic.
We additionally report F1 scores at selected false-positive rates to examine performance at specific decision thresholds.
We report both aggregate results across attacks in this section, and detailed per-attack results in~\cref{appendix:kitsune},~\cref{appendix:hv}, and~\cref{appendix:planter}.
Detailed descriptions of the experimental testbed, datasets, training procedure, and metrics are provided in~\cref{appendix:eval-meth}.

\subsection{Detection Performance vs. Kitsune}
\label{subsec:eval-detection-kitsune}

This section compares \systemname-K against Kitsune~\cite{mirsky2018kitsune} across 15 attacks from both the original Kitsune evaluation and the CIC-IDS-2018 dataset.
Figure~\ref{fig:auc-kitnet-violin} summarizes the distribution of AUC scores across attacks for each sampling rate using violin plots.
The central marker indicates the median AUC, the upper and lower bounds correspond to the third and first quartiles, respectively, and the violin width represents the density of attacks achieving a given AUC value.
Individual attack results are presented in Appendix~\ref{appendix:kitsune}.

Both systems achieve strong detection performance in the unsampled setting (1:1).
However, their behavior diverges sharply as sampling increases.
Kitsune's median AUC drops rapidly under pre-feature packet sampling, and the distribution progressively concentrates around random guessing.
Under aggressive pre-feature sampling, most attacks approach or fall below chance-level ranking, indicating that the statistical features required by Kitsune can no longer be reconstructed from the sparse packet stream presented to the detector.

In contrast, \systemname-K remains substantially robust to sampling.
Although detection effectiveness varies across attacks, the median AUC remains consistently higher than that of Kitsune across all evaluated sampling rates, and the distribution retains a significant fraction of attacks with strong detection performance even under aggressive traffic reduction.
These results suggest that \textit{computing approximate features over all traffic is more effective than computing exact features from a heavily sampled packet stream}.

The aggregate trends visible in Figure~\ref{fig:auc-kitnet-violin} are consistent with the detailed per-attack analysis presented in Appendix~\ref{appendix:kitsune}.
At a sampling rate of 1:1024, \systemname-K achieves AUC $>$ 0.5 on 14 of the 15 attack traces, compared with three for Kitsune, and retains AUC $>$ 0.8 on 12 traces.

These results show that, for the same underlying detector, preserving packet contributions before sampling matters more than computing exact features from a sampled packet stream.

\subsection{Detection Performance vs. HyperVision}
\label{subsec:eval-detection-hv}

This section compares \systemname-H against HyperVision~\cite{fu2023hypervision} across the full set of 69 attacks from the original HyperVision evaluation.
Figure~\ref{fig:auc-hv-violin} summarizes the distribution of AUC scores across attacks for each sampling rate.
Individual attack results are presented in Appendix~\ref{appendix:hv}.

As expected, HyperVision generally achieves higher detection performance in the unsampled (1:1) setting.
However, its effectiveness degrades rapidly as sampling increases.
The median AUC drops sharply, and the distribution progressively concentrates around random guessing, indicating that pre-feature packet sampling removes much of the structural and temporal information required by the detector.

In contrast, \systemname-H exhibits substantially greater robustness to sampling.
Although detection performance varies considerably across attacks, the median AUC remains relatively stable as sampling increases, and the distribution does not exhibit the systematic collapse observed for HyperVision.
This trend mirrors the results observed for \systemname-K in Figure~\ref{fig:auc-kitnet-violin}, despite the two systems employing fundamentally different detection pipelines, suggesting that the benefits arise primarily from preserving visibility prior to traffic reduction rather than from the specific ML model.

This aggregate view is consistent with the per-attack results presented in Appendix~\ref{appendix:hv}.
At 1:4096, \systemname-H achieves AUC $>$ 0.5 on 41 of the 69 attack traces and AUC $>$ 0.7 on 24, whereas HyperVision achieves AUC $<$ 0.5 for the majority of attacks, and no AUC above 0.7.

\subsection{Detection Performance vs. Planter}
\label{subsec:eval-detection-planter}

This section compares \systemname-K against Planter~\cite{zheng2024planter}, a representative in-switch ML framework.
Unlike Kitsune and HyperVision, which perform feature computation and ML inference in software, Planter executes both operations directly in the data plane.
We evaluate an Isolation Forest configuration and an autoencoder (AE) baseline configured following Planter's original evaluation, focusing on anomaly detection without attack-specific training examples.
The AE baseline uses a modified KitNET autoencoder that matches Planter's base configuration. 
\cref{appendix:planter} details this adaptation.
Figure~\ref{fig:auc-planter-violin} summarizes the resulting AUC distributions across the evaluated attacks.

The evaluated configurations trade feature and model expressiveness for per-packet inference in the switch.
Our comparison measures the resulting system-level detection performance.
Whereas Kitsune and HyperVision suffer from visibility loss due to pre-feature packet sampling, Planter avoids sampling entirely by executing inference in the switch.
However, the computational constraints of programmable switches severely restrict model complexity, forcing Planter to rely on substantially simpler learning algorithms than the unsupervised models employed by \systemname.

As shown in Figure~\ref{fig:auc-planter-violin}, \systemname-K consistently achieves higher detection effectiveness than the Planter configurations.
While the performance of the Planter models varies across attacks, their median AUC remains significantly below that of \systemname-K.
This result suggests that preserving expressive ML models in software provides greater benefits than executing simplified inference directly in the data plane.

The aggregate results are consistent with the per-attack analysis presented in Appendix~\ref{appendix:planter}.
At a post-feature sampling rate of 1:1024, \systemname-K achieves higher AUC than the evaluated IF and AE baselines on 62 and 69 of the 84 attack traces, respectively.
These paired comparisons show that exporting fewer, richer feature records to software can outperform the evaluated in-switch configurations, even though those configurations process every packet.

In-switch inference retains an advantage in decision latency.
\systemname instead uses switch resources to construct richer inputs for software detection.

\subsection{Feasibility and System Efficiency}
\label{subsec:system-implications}

We now evaluate switch resource usage, flow-state capacity, communication overhead, throughput, and deployment cost.


\noindent
\textbf{Data plane resource usage.}
\systemname consumes a significant fraction of switch resources, reflecting the migration of feature computation from software servers into the data plane.
On Tofino 2, \systemname-K uses 26.9\% of SRAM and 72.4\% of meter ALUs, while \systemname-H uses 46.2\% and 86.3\%, respectively.
Their computations span 95\% and 90\% of pipeline stages.
Despite this overhead, the design fits within the resource budget of current programmable switches while maintaining L2 forwarding functionality and line-rate operation.
Detailed compiler reports and resource breakdowns are provided in Appendix~\ref{appendix:resource-usage}.

\begin{figure*}[t]
\centering
	\begin{minipage}[t]{.5\textwidth}
		\centering
        \includegraphics[width=.95\linewidth]{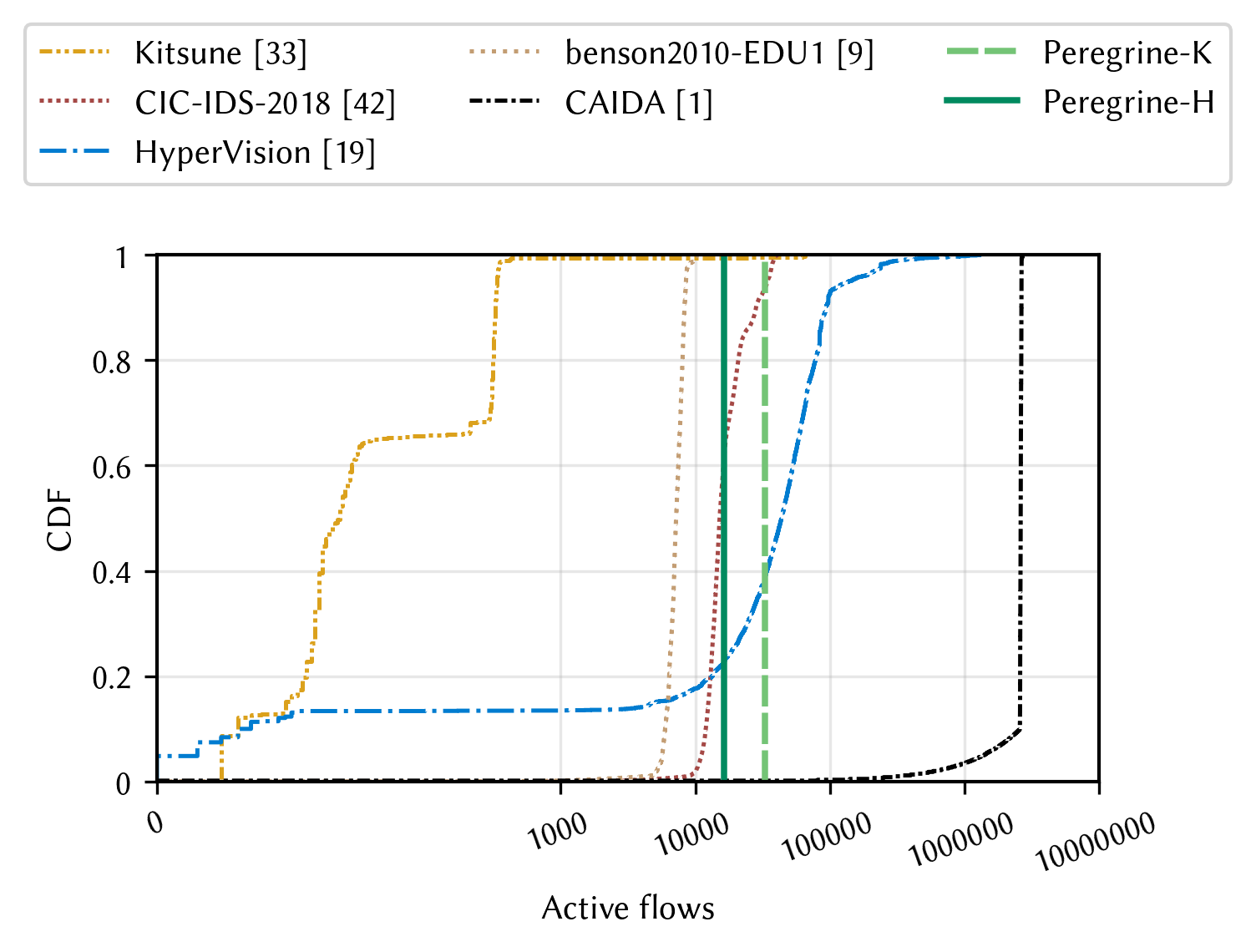}		\captionsetup{width=.8\linewidth}
		\captionof{figure}{Active flows per second (CDF) for the tested datasets, along with a sample trace from a CAIDA dataset. Vertical lines represent the flow storage capacity of \systemname-K and \systemname-H.}
		\label{fig:active-flow-cdf}
	\end{minipage}%
	\begin{minipage}[t]{.5\textwidth}
		\centering
		\includegraphics[width=.95\linewidth]{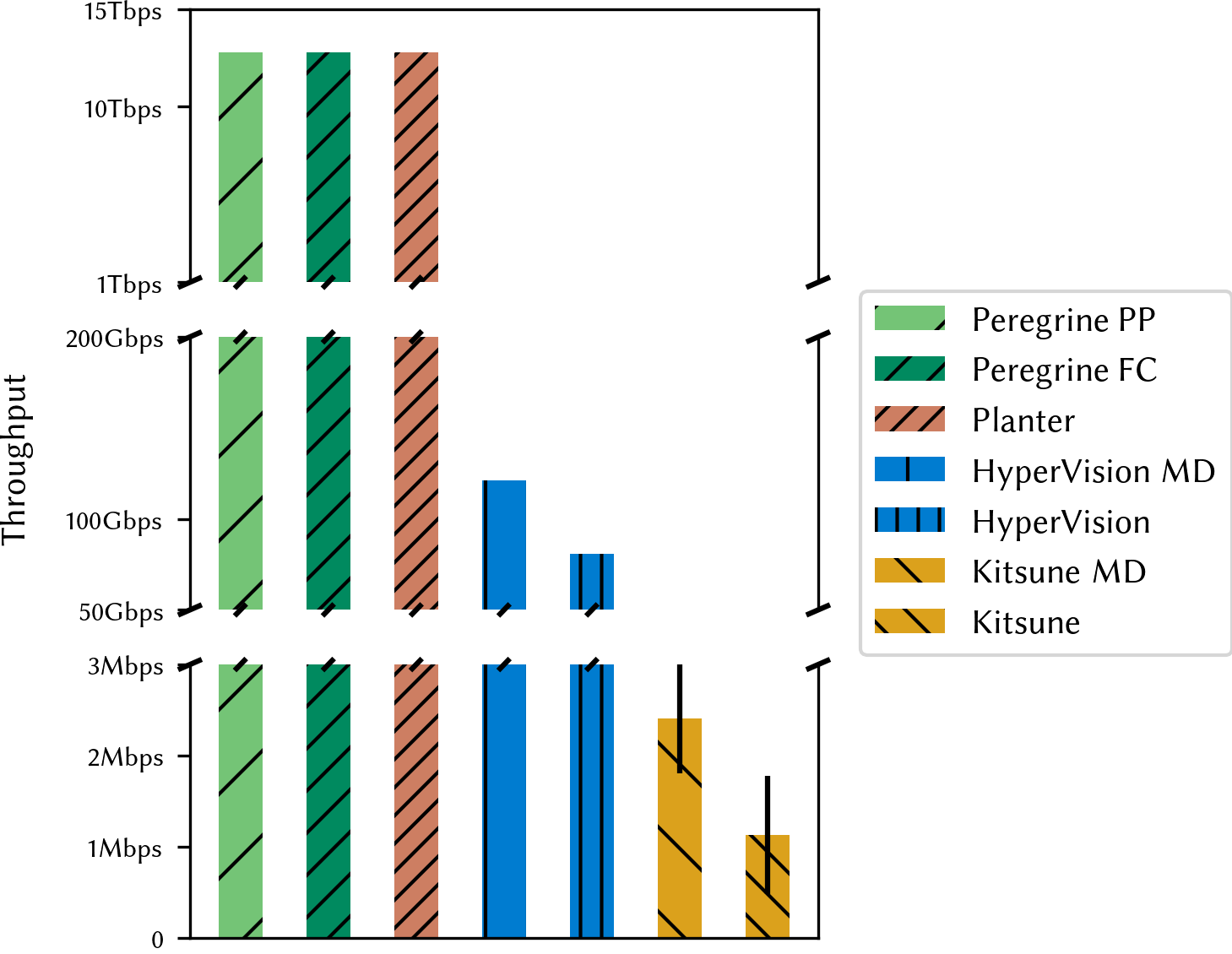}
		\captionsetup{width=.8\linewidth}
		\captionof{figure}{Pipeline performance of \systemname~vs. competitor systems.}
		\label{fig:pipeline_perf}
	\end{minipage}
\end{figure*}

\noindent
\textbf{Flow-state scalability.}
\systemname stores feature atoms in switch SRAM, whose total capacity is approximately 200 Mb (25 MB) on a Tofino 2.
The memory footprint per active flow is 156 bytes for \systemname-K and 60 bytes for \systemname-H.
However, scalability is constrained not only by aggregate SRAM capacity but also by the Tofino architecture: feature atoms must be distributed across multiple equal-sized stateful registers accessed in specific pipeline stages, each with an independent memory budget.
Consequently, the total number of registers required for both feature atom storage and feature computation operations determines the provisioned state capacity: 32k entries for \systemname-K and 16k for \systemname-H.
These entry counts are upper bounds on the number of independently tracked flows; hash collisions can reduce the effective capacity, while state lifetime determines the resident working set.
Our implementation currently prioritizes demonstrating the detection benefits of sampling placement rather than maximizing flow-state capacity.

Figure~\ref{fig:active-flow-cdf} shows the distribution of flow counts across the evaluated and operational traces.
The provisioned capacity supports a relevant deployment regime: high-bandwidth monitoring with bounded flow populations, as found in AI training fabrics~\cite{Qian2024AlibabaHPN} and rack-level measurements of data center traffic~\cite{Benson2010Wild}.
These environments motivate our design: high traffic volume need not imply high flow cardinality.
The HyperVision traces and high-cardinality Internet traces such as CAIDA include flow counts exceeding the provisioned capacities.
Extending support to these workloads requires additional mechanisms for state reduction or approximation, which we discuss in Section~\ref{sec:discussion}.

\noindent
\textbf{Communication overhead.}
A secondary benefit of \systemname is the reduction in switch-to-server communication when compared to ML-based detectors~\cite{mirsky2018kitsune,fu2023hypervision}.
Existing detectors typically export sampled packets to a server for feature computation and inference.
In contrast, \systemname performs feature computation in the data plane and exports only compact feature records.

To quantify this effect, we measured the average packet size across the evaluated datasets and operational traces.
The average packet size ranges from 381 B (CIC-IDS-2018~\cite{sharafaldin2018toward}) to 1131 B (HyperVision~\cite{fu2023hypervision}), with intermediate values of 644 B (benson2010-UNI1~\cite{Benson2010Wild}), 869 B (CAIDA~\cite{passive_2019_pcap}), and 890 B (Kitsune~\cite{mirsky2018kitsune}).
Since \systemname-K and \systemname-H export fixed-size feature records of 156 B and 60 B, respectively, these packet sizes correspond to communication reductions ranging from 2.4$\times$ to 7.3$\times$ for \systemname-K and from 6.4$\times$ to 18.9$\times$ for \systemname-H relative to packet sampling approaches.
This gain is conceptually similar to operational monitoring systems such as sFlow/NetFlow, which reduce monitoring overhead by exporting compact summaries, rather than full traffic streams.

\noindent
\textbf{Pipeline performance across system components.}
Figure~\ref{fig:pipeline_perf} compares the throughput of the Feature Computation (FC) and ML-based Detection (MD) components against prior systems, including Kitsune, HyperVision, and Planter.
Using the snake-topology setup described in Appendix~\ref{appendix:eval-meth}, \systemname sustains packet processing and feature computation at aggregate traffic rates up to 12.8 Tbps, matching the Tofino 2 ASIC's aggregate switching capacity and demonstrating Terabit-scale feature computation on real hardware.
The MD component's throughput depends on the underlying ML model and its implementation.
\systemname-K's MD component (based on Kitsune) achieves throughputs in the Mbps range\footnote{The original Kitsune runtime evaluation uses a C++ implementation~\cite{mirsky2018kitsune}.
We use the authors' publicly available Python implementation, which they explicitly
describe as not optimized for speed.
Our measurements therefore characterize this implementation rather than the maximum
throughput achievable by KitNET.}, while \systemname-H's MD component (based on HyperVision) reaches approximately 120 Gbps.
By offloading feature computation to the switch, the server is relieved from a major processing burden (details in~\cref{appendix:runtime-perf}) and can devote more processing capacity to ML inference.
For HyperVision, throughput increases from approximately 80 Gbps to 120 Gbps, a 1.5$\times$ improvement.

The measured MD throughputs illustrate the scale of the throughput mismatch between switches and ML detectors.
A sustainable execution requires the exported feature-record rate to remain within the backend's processing capacity.
Lower-throughput detectors therefore require aggressive record reduction or additional backend capacity.
In particular, the evaluated Python-based
\systemname-K backend requires a greater reduction than 1:1024, in order
to sustain Terabit-scale input without additional processing capacity.
The optimized KitNET implementation~\cite{mirsky2018kitsune} would, however, reduce the required downsampling relative to our Python backend.
These results demonstrate that feature computation can sustain Terabit-scale traffic, while end-to-end throughput depends on matching record emission to the software backend.
Crucially, this allows traffic reduction to occur after feature computation rather than before it, preserving packet contributions to the features used for detection.

Future advances in ML acceleration may reduce the amount of traffic reduction required and eventually eliminate it for some detectors.
However, whenever downstream ML processing remains slower than the monitored line rate, computing features before reducing the exported record stream preserves information that pre-feature packet sampling would discard.
\systemname is therefore complementary to advances in ML acceleration rather than a replacement for them.

\noindent
\textbf{Cost and energy efficiency.}
By moving per-packet feature computation into the programmable switch, \systemname reduces the server resources required to monitor Terabit-scale traffic without discarding packets before feature extraction.
Using the methodology described in Appendix~\ref{appendix:efficiency}, we estimate the number of servers required by existing systems to process equivalent traffic volumes without sampling and compare their aggregate power consumption and deployment cost against a \systemname deployment.
Figure~19 presents these model-based estimates, using detector throughput measurements and published hardware cost and power figures.
Under the assumed per-server throughput, the cost and power consumption of unsampled software-based detection grow approximately linearly with monitored traffic volume, as packet processing, feature extraction, and inference require an increasingly higher number of servers.
In contrast, \systemname performs packet processing and feature computation within the switch's capacity, while the software inference tier is provisioned according to the exported feature-record rate. The analysis illustrates the potential savings from replacing software-based per-packet processing with in-switch feature computation and post-feature downsampling.
Detailed assumptions and cost and energy estimates are provided in Appendix~\ref{appendix:efficiency}.

%% file: sections/discussion.tex
\section{Discussion and Limitations}
\label{sec:discussion}


\noindent
\textbf{Deployment trade-offs.}
Maintaining feature state and computing features at line rate consumes a substantial fraction of switch resources.
\systemname is therefore best suited to a dedicated monitoring switch or helper switch on the forwarding path.
This design exchanges switch resources for reduced server-side processing.
Software inference can evolve without changing the data plane when feature requirements remain unchanged.
Changes to the extracted features may require modifying and redeploying the P4 pipeline.
Importantly, packet forwarding remains active (L2 forwarding) and line-rate throughput is preserved throughout our implementation.

\noindent
\textbf{Scalability.}
State capacity and backend throughput impose distinct limits (\cref{subsec:system-implications}).
Our implementations provision approximately 32k state entries for \systemname-K and 16k for \systemname-H, but effective flow capacity depends on collisions and state lifetime.
These capacities suit bounded-flow deployments but do not cover all workloads in Figure~\ref{fig:active-flow-cdf}.
More compact state representations or external-memory architectures such as TEA~\cite{kim2020tea} could expand this range, with corresponding accuracy or access-latency trade-offs.
Distributing monitoring across multiple switches could also increase aggregate capacity, provided that traffic placement preserves the flow and cross-flow relationships required by the detector.

\noindent
\textbf{Detection latency.}
\systemname produces decisions from exported feature records.
In \systemname-K, record emission is controlled by a global packet-count epoch; in \systemname-H, it depends on timeout detection and the configured export interval.
End-to-end detection latency also includes record transport, backend queueing, and inference.
In-switch ML~\cite{xiong2019switches,busse2019pforest,zhou2023efficient,yan2024brain,zheng2024planter,zhang2025quark} can produce per-packet decisions with lower latency, whereas \systemname prioritizes richer features and software inference, achieving higher AUC than the evaluated in-switch configurations (~\cref{subsec:eval-detection-planter}).
As stated in \cref{sec:threatmodel}, the system targets monitoring and detection rather than guaranteed prevention before an attack's first damaging packet.

\noindent
\textbf{Alternative design points.}
Exporting feature atoms periodically and completing feature computation in software would also preserve packet contributions before traffic reduction.
This alternative reduces switch computation at the expense of additional per-record processing on the server.
Our implementation completes these computations in the switch.
The preferred partition depends on the available switch resources and backend capacity.
Another alternative entails performing flow sampling instead of packet sampling. 
Flow sampling retains all packets of selected flows, preserving their within-flow statistics while excluding other flows and potentially altering cross-flow statistics and graph structure.
We therefore expect it to improve on packet sampling for detectors dominated by within-flow features, while trading traffic coverage for feature fidelity.
Post-feature sampling instead allows every packet to contribute to feature state, even when no corresponding record is exported.

%% file: sections/conclusion_future_work.tex
\section{Conclusion}
\label{sec:conclusion}

Machine learning-based malicious traffic detection at Terabit scale is constrained not only by model accuracy, but also by the architecture used to deliver traffic features to the detector.
Aggressive pre-feature packet sampling can discard the behavioral signals required for effective detection.

This paper shows that the key scalability question is not only how much to sample, but also \textit{where} sampling occurs.
By computing ML features over all traffic at line rate in the programmable switch data plane and applying downsampling only after feature computation, \systemname sustains Terabit-scale feature computation and preserves detection effectiveness under aggressive traffic reduction across two distinct feature-extraction pipelines.
Our evaluation across 84 attack traces demonstrates that approximate features computed over all packets can outperform exact features computed from sampled packets.
More broadly, \systemname shows how programmable switches can preserve the information needed by ML detectors, making sampling placement a central design consideration for high-speed security and network analytics.

%% file: sections/acknowledgements.tex
\section*{Acknowledgements}

This work was supported by the European Union (ACES project, 101093126) and by national FCT funds (Myriarch project, 2022.09325.PTDC).
João Romeiras Amado was supported by the FCT scholarship 2020.05965.BD.

%% file: sections/appendix-implementation.tex
\subsection{Implementation}
\label{appendix:implementation}

The \systemname prototype consists of a programmable data plane implementation and a control plane software stack that together realize the architecture described in Section~\ref{sec:design}.
This section focuses on the practical challenges of implementing \systemname on real switching hardware, and on the techniques required to map the abstract feature-computation model onto a constrained PISA pipeline.
The goal is not to replicate software feature computation exactly, but to preserve the behavioral signals required for ML-based detection while operating at line rate.

The implementation of both \systemname-K and \systemname-H consists of a few thousand lines of P4 code for the data plane, and a few thousand lines of C++ for the control plane software running on a middlebox server.
The switch data plane executes a P4 program that implements packet processing and feature computation, while the server processes feature records and performs ML-based detection.
For \systemname-K, the ML backend is implemented in Python using KitNET~\cite{mirsky2018kitsune}, whereas for \systemname-H, the graph processing and ML pipelines are written in C++ and derived from HyperVision~\cite{fu2023hypervision}.
We make the P4 implementations of \systemname publicly available~\cite{anonarchrepo}.

The data plane implementation targets the Tofino Native Architecture (TNA)~\cite{tna}, a realization of the PISA model used by Intel’s Tofino switching ASICs.
We support both Tofino 1 (TNA) and Tofino 2 (T2NA), which differ primarily in available pipeline resources and total throughput.
\systemname-K includes implementations for both generations, while \systemname-H targets Tofino 2 due to its higher state and computation requirements.

To conform to the physical constraints of these platforms, several feature-computation operations described in \cref{subsec:peregrine-data-plane} rely on approximations and non-trivial implementation strategies.
We describe these techniques next.

\subsubsection{Implementation: \systemname-K}
\label{sec:impl-peregrine-kitnet}

This subsection showcases how \systemname-K realizes  statistical feature computation and decay in a streaming fashion, within the constraints of a PISA pipeline.

\begin{table}[t]
	\caption{Statistical Features calculated in the data plane.}
	\label{tab:primitives}
	\small
	\centering
	\begin{tabular}{@{}lcc@{}}
		\hline
		\textbf{Statistics} & \textbf{Notation} & \textbf{Calculation} \\ \hline
		Weight & $w$ & $w$ \\ \hline
		Mean & $\mu$ & $\frac{LS}{w}$ \\ \hline
		Std. Deviation & $\sigma_{S_{i}}$ & $\sqrt{\mid\frac{SS}{w} - (\frac{LS}{w})^{2}\mid}$ \\ \hline
		Magnitude\textbf{*} & $\mid\mid S_{i},S_{j} \mid\mid$ & $\sqrt{\mu^{2}_{S_{i}} + \mu^{2}_{S_{j}}}$ \\ \hline
		Radius\textbf{*} & $R_{S_{i},S_{j}}$ &  $\sqrt{(\sigma^{2}_{S_{i}})^{2} + (\sigma^{2}_{S_{j}})^{2}}$ \\ \hline
		Approx. Covariance\textbf{*} & $Cov_{S_{i},S{j}}$ & $\frac{SR_{ij}}{w_{i} + w_{j}}$ \\ \hline
		Pearson Corr. Coeff.\textbf{*} & $PCC_{S_{i},S_{j}}$ & $\frac{Cov_{S_{i},S_{j}}}{\sigma_{S_{i}}\sigma_{S_{j}}}$ \\ \hline
	\end{tabular}
	\\
	\vspace{0.4em}
	\textbf{* Bidirectional statistics.} \\
	\textit{LS = linear sum of the packet sizes} \\
	\textit{SS = squared sum of the packet sizes} \\
	\textit{$SR_{ij}$ = sum of residual products ($Res_{i,j}$) for streams $i$ and $j$}
	\vspace{-1em}
\end{table}

\noindent
\textbf{Approximating Arithmetic.}
\systemname-K's feature atoms and derived statistics require arithmetic operations that are not natively supported by the simple ALUs available in PISA pipeline stages, including multiplication, division, exponentiation, and square root.
To implement these operations at line rate, we rely on a set of approximation techniques that trade numerical precision for pipeline compatibility.

The first technique approximates multiplication and division using bit-shift operations.
The second operand (e.g., a divisor) is rounded to the nearest power of two, and the corresponding operation is implemented using a logical shift.
The appropriate rounding action is selected using ternary match tables.
The second technique leverages Tofino math units---specialized hardware blocks exposed via P4 externs---that support arithmetic with one constant operand.
These units are used to implement operations such as scaled division and approximate exponentiation.
Finally, exponentiation and square root are implemented using low-precision approximations provided by the same math units.
While these approximations introduce bounded error, they preserve the qualitative behavior of the features and are sufficient for downstream ML inference, as validated in ~\cref{sec:evaluation}.

\begin{figure*}[t]
	\centering
	\subfigure[Handling multiple decay values.]{\includegraphics[width=.49\textwidth]{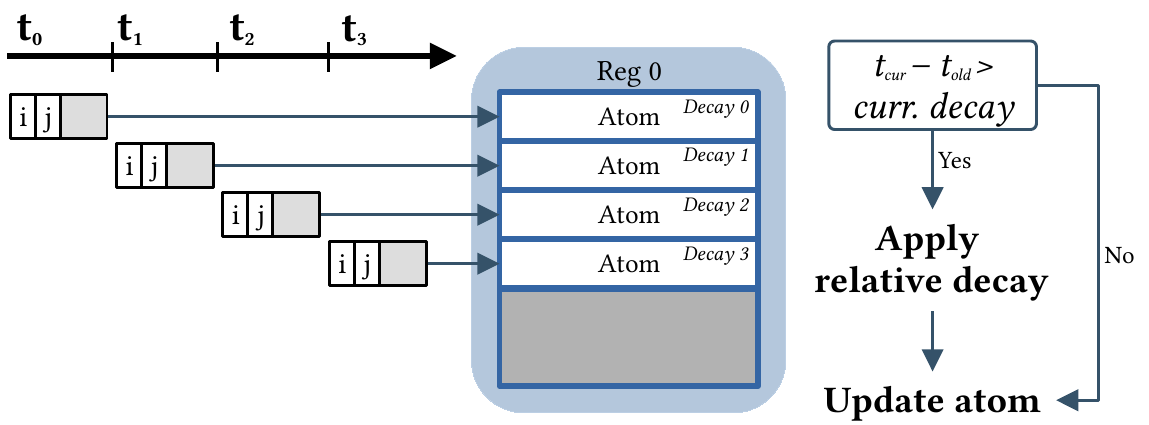}}
	\subfigure[Tracking bidirectional traffic.]{\includegraphics[width=.49\textwidth]{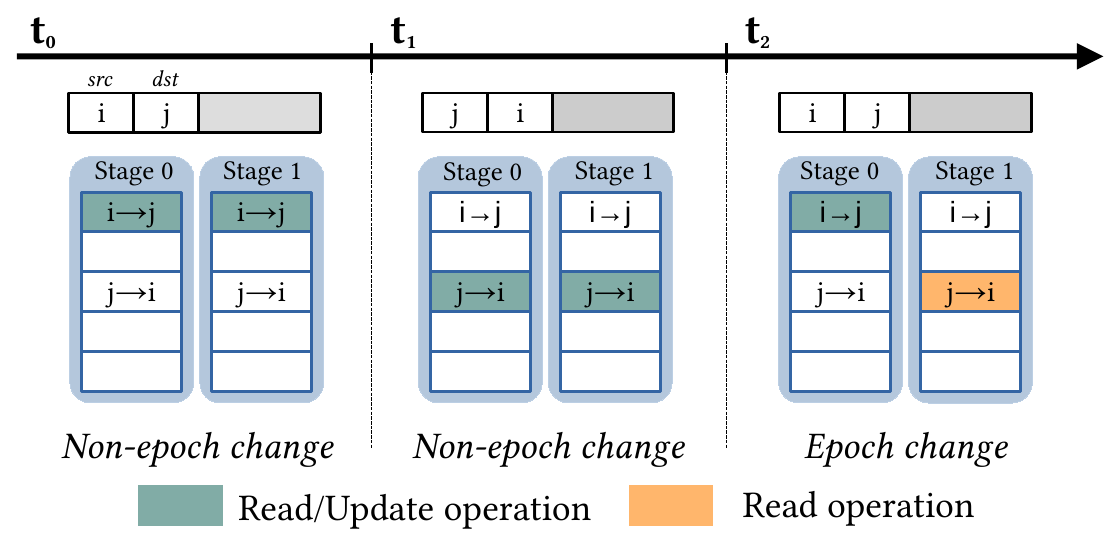}}
	\caption{\systemname-K: Data Plane Implementation.}
	\label{fig:peregrine-k-impl}
\end{figure*}

\noindent
\textbf{Handling Multiple Decay Factors.}
\systemname-K applies multiple temporal decay factors to each feature atom to capture traffic dynamics at different time scales.
Specifically, four decay intervals are supported (100ms, 1s, 10s, 60s), corresponding to decay factors $\lambda=(10,1,\frac{1}{10},\frac{1}{60})$ (\cref{subsec:peregrine-k-data-plane}).

Implementing all four decays simultaneously is infeasible due to the single register access per packet constraint.
Rather than replicating feature atoms across multiple pipeline stages---an approach that would incur prohibitive resource usage---\systemname-K updates one decay instance per packet, rotating among decay factors across successive packets (Figure~\ref{fig:peregrine-k-impl}(a)).
This approach sacrifices exact temporal alignment in favor of efficiency.
In cases where inter-arrival times significantly exceed the selected decay interval, decay is applied iteratively across packets until convergence.
This design reflects a deliberate trade-off: precise decay timing is less critical than ensuring that all packets contribute to the feature state.

\noindent
\textbf{Tracking Bidirectional Traffic.}
Computing bidirectional statistics (Table \ref{tab:primitives}) requires correlating feature atoms across both directions of a flow.
Because a register array can be accessed only once per packet, storing both directions in a single stage is infeasible.

\systemname-K addresses this by duplicating correlated feature atoms across two pipeline stages.
During normal operation, each stage updates the atom corresponding to the packet's direction.
Once per epoch, the second stage reads the inverse direction to compute bidirectional statistics (Figure~\ref{fig:peregrine-k-impl}(b)).
This design introduces a temporary inconsistency between duplicated atoms during epoch transitions.
The inconsistency is bounded and resolved periodically by synchronizing atom values, and does not affect detection accuracy in practice (its impact is assessed through the detection evaluation in \cref{sec:evaluation}).

\subsubsection{Implementation: \systemname-K for Tofino 1}
\label{sec:impl-peregrine-kitsune-tna}

\begin{figure*}[t]
	\centering
	\includegraphics[width=1\linewidth]{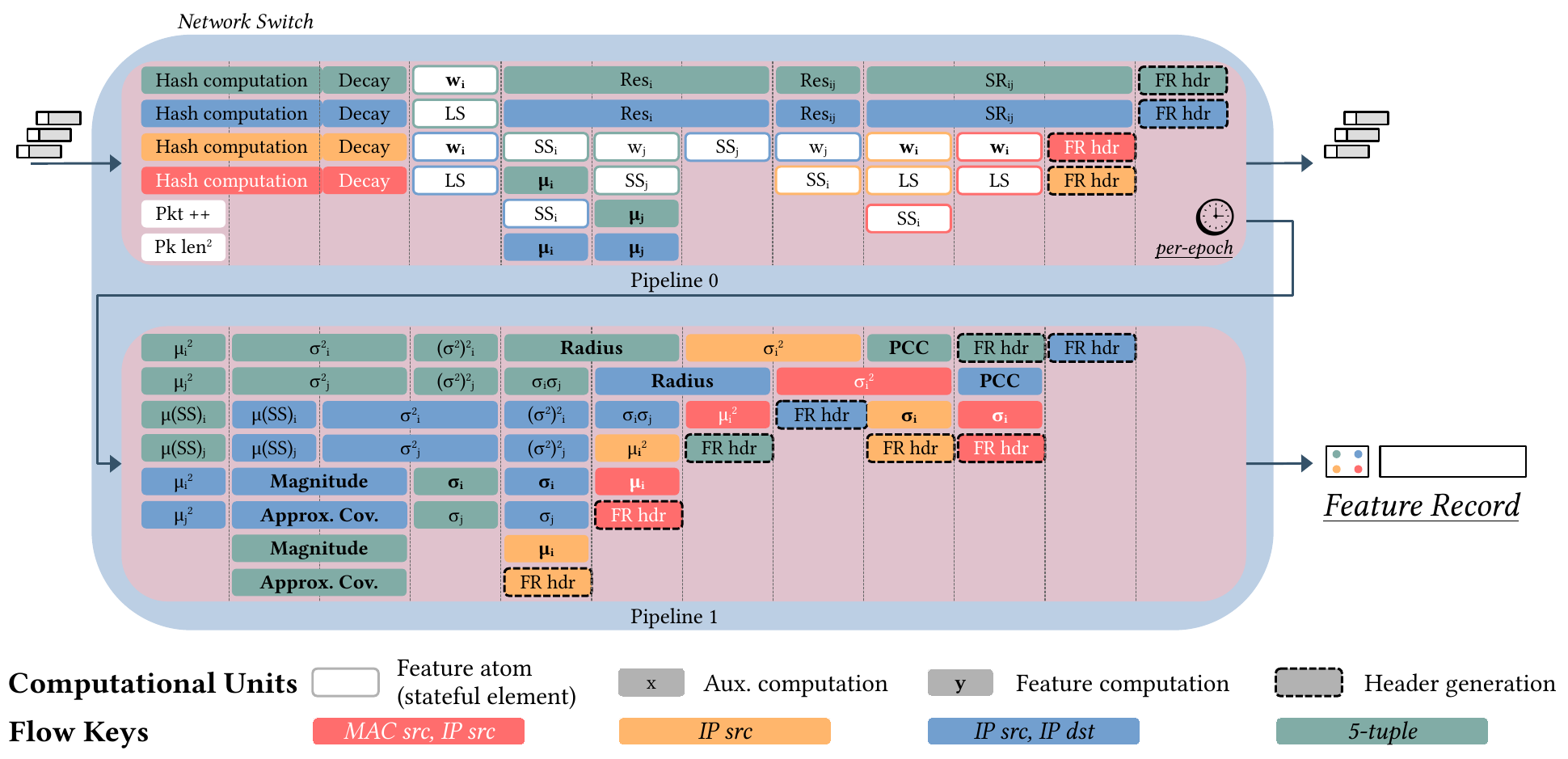}
	\vspace{-1em}
	\captionof{figure}{\systemname-K: Data plane implementation of the feature computation module for the Tofino 1 (TNA) architecture, with recirculation across two pipelines. Each column inside a pipeline represents a stage. Several computations are performed per-stage across all flow keys.}    
	\label{fig:peregrine-arch-tna1}
\end{figure*}

The Tofino 2 (T2NA) architecture features a higher number of stages per pipeline compared to the previous generation of the Tofino switch architecture (TNA).
As such, \systemname-K's implementation for T2NA is able to compute the entire set of features for all flow keys on a single pipeline.
However, the number and complexity of the features computed by \systemname-K precludes its deployment on a single pipeline of a TNA-based switch.

To overcome this issue and enable the use of \systemname-K using the first generation of Tofino chips, we leverage the packet recirculation primitive of this switch architecture to forward only selected packets to another pipeline in the switch.
This primitive \textit{virtually} extends the number of stages for the additional processing required by \systemname-K, as shown in Figure~\ref{fig:peregrine-arch-tna1}.
Using two pipelines on TNA introduces overhead, as the recirculation ports only support a relatively small portion of the switch's overall throughput.
However, in \systemname-K, recirculation needs only be performed once per epoch to support the computation of the more complex features and subsequent record generation.
Notably, the feature atoms \textit{are executed in the first pipeline}, enabling their updates for each packet traversing the switch---in other words, we observe and maintain the state of every packet, which is fundamental to guarantee a high detection rate.
On T2NA, an operator's configuration of the epoch value is not restrained by the limitations on the overall switch throughput as recirculation is not necessary, being instead only dependent on the throughput handling capabilities of the active classifier running on the middlebox server.

As can be observed in the high-level illustration of the switch's pipelines in Figure \ref{fig:peregrine-arch-tna1}, a number of auxiliary computations are performed in the data plane to calculate the more complex features. 
These computations are strategically split between both pipelines in order to optimize pipeline placement.
The values obtained from the auxiliary computations performed in the first pipeline are carried onto the second pipeline using internal bridge headers and subsequently used in the remaining feature computation operations.
The complete list of statistics computed in the data plane of \systemname-K is included in Table~\ref{tab:primitives}.

\noindent
\textbf{Computations in detail.}
We conclude the \systemname-K implementation section with a detailed description of the individual computations, illustrated in Figure~\ref{fig:peregrine-arch-tna1} for Tofino 1.
The same computations are used on Tofino 2, but fit within a single switch pipeline and therefore require no packet recirculation.
For each computation, we indicate the total number of required processing stages in square brackets.

\noindent
\textit{Hash Computation [2 stages]:} Usually a single-stage process, \systemname-K's hash computation requires an extra step performed in a second stage.
Since, for a given flow key, the 4 instances of each flow atom---corresponding to the 4 decay constants---are stored in the same register, each occupying a quarter of its available memory, we must add a constant value to the obtained hash, representing the register index value that marks the initial position for the active decay factor.

\noindent
\textit{Pkt$++$ [1 stage]:} Global packet counter, used to keep track of each epoch's state.

\noindent
\textit{Residue ($Res_i$) [3 stages]:} \systemname-K tracks a residue value for each flow direction, with both values stored in a single register position using a P4 struct.
A hash-based check is performed to track the previous/current flow direction and update the stored values in the correct struct positions accordingly.
This operation encompasses: (1) Residue value calculation (subtraction of the packet length and mean); (2) Flow direction check; (3) Stored residue values' update.  

\noindent
\textit{Sum of residual products ($SR_{ij}$) [3 stages]:} Calculated as 64-bit values, the sums of residual products are obtained through a process encompassing three pipeline stages.
Due to architecture limitations that restrict register actions on 64-bit values, the required calculations are performed using a sequence of 32-bit manipulations: (1) Sum of the lower 32-bits; (2) Carry value calculation; (3) Sum of the higher 32-bits.

\noindent
\textit{Mean ($\mu_{i}$) [1 stage]:} Calculated through a right-shift division between the linear sum of the number of packet bytes and the number of packets (both obtained as feature atoms).

\noindent
\textit{Mean of Squared Sum ($\mu(SS)_{i}$) [1 stage]:} Calculated through a right-shift division between the squared sum of the number of packet bytes and the number of packets (both obtained as feature atoms).

\noindent
\textit{Squared Mean ($\mu^{2}_{i}$) [1 stage]:} Obtained through a square calculation (math unit approximation) of the mean for a given flow.

\noindent
\textit{Variance ($\sigma^{2}_{i}$) [2 stages]:} Obtained through (1) Subtraction of the mean (squared sum) and the squared mean; (2) Square root calculation (math unit approximation) of the previously obtained value.

\noindent
\textit{Standard Deviation ($\sigma_{i}$) [1 stage]:} Obtained through a square root calculation (math unit approximation) of the variance.

\noindent
\textit{Magnitude [2 stages]:} Obtained through (1) Addition of the squared mean for both flow directions; (2) Square root calculation (math unit approximation) of the previously obtained value.

\noindent
\textit{Radius [2 stages]:} Obtained through (1) Addition of the squared variance for both flow directions: (2) Square root calculation (math unit approximation) of the previously obtained value.

\noindent
\textit{Approximate Covariance (Approx. Cov.) [2 stages]:} Obtained through (1) Addition of the packet weight for both flow directions; (2) Right-shift division for the sum of residual products and added weights.

\noindent
\textit{Pearson Correlation Coefficient (PCC) [1 stage]:} Calculated through a right-shift division between the approximate covariance and the product of the standard deviation for both flow directions.

\subsubsection{Implementation: \systemname-H}
\label{sec:impl-peregrine-hypervision}

\systemname-H stresses different aspects of the \systemname architecture than \systemname-K, particularly the maintenance of per-flow feature distributions and timeout-driven record generation.
Unlike streaming statistical aggregates, \systemname-H requires the data plane to (i) update active flows on packet arrival and (ii) periodically identify inactive flows whose feature records must be emitted to the control plane.
Reconciling these requirements under the strict execution constraints of programmable switches requires careful design.

\noindent
\textbf{Flow update and timeout detection.}
\systemname-H determines when a flow should be exported to the control plane based on a configurable inactivity timeout.
In addition to updating the feature state of the flow associated with each incoming packet, the switch must continuously check whether other active flows have exceeded the timeout threshold.

\begin{figure}[t]
	\centering
	\includegraphics[width=0.5\textwidth]{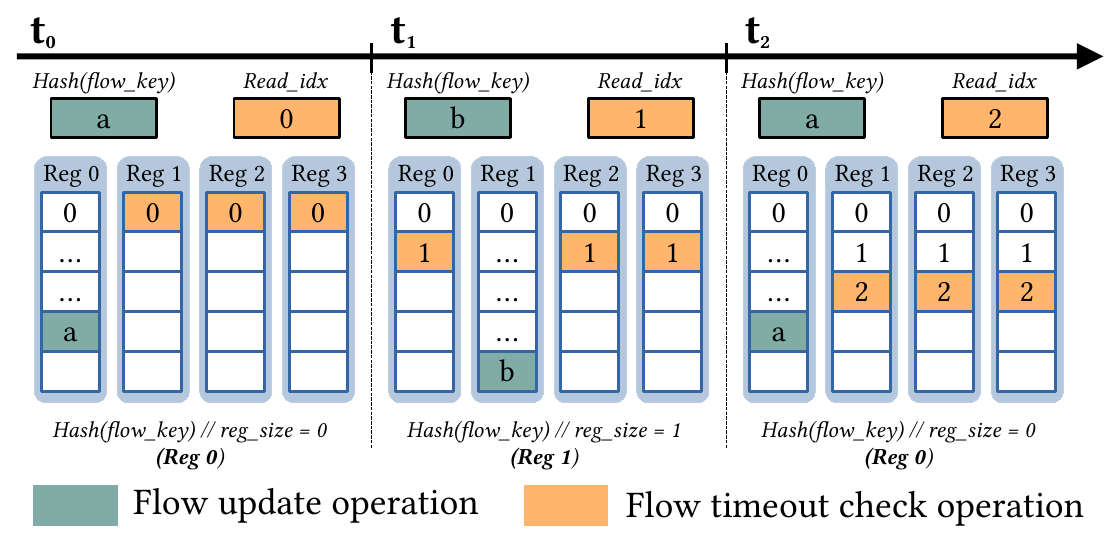}
	\caption{\systemname-H. Flow update / Timeout check.}
	\label{fig:flow-timeout-check}
\end{figure}

However, programmable switches allow only a single access to each stateful memory element per packet, preventing a naive design that simultaneously updates one flow and checks timeouts for another.
To address this limitation, \systemname-H partitions the per-flow state across multiple independent register banks, each holding a disjoint subset of the flow table (Figure~\ref{fig:flow-timeout-check}).
For each packet, the flow corresponding to the packet is updated in one register bank, while timeout checks are performed on entries in the remaining banks.

The specific register bank to update is selected via a hash of the flow key, ensuring uniform distribution of flows across banks.
A global timeout index is advanced on every packet, allowing all flow entries to be checked incrementally over time.
This design guarantees that every flow is eventually examined for timeout expiration without violating per-packet access constraints.

\noindent
\textbf{Flow distribution fitting using feature bins.}
In addition to scalar per-flow features, \systemname-H maintains compact per-flow feature distributions for packet inter-arrival times and packet lengths.
For each such feature, the data plane defines a fixed number of bins with operator-configured ranges.
Upon packet arrival, the appropriate bin counter is incremented based on the observed value.
When a flow expires or an epoch ends, the feature record exported to the control plane includes the full set of bin counters, enabling the ML backend to reason about traffic distributions rather than simple aggregates.
The same register-partitioning mechanism is applied to bin counters, allowing distribution updates and timeout checks to proceed concurrently at line rate.

\noindent
\textbf{Discussion.}
This implementation preserves the same architectural contract as \systemname-K: all packets contribute to the feature state in the data plane, while only compact feature records are subject to downsampling before ML inference.
Despite the different complexity of \systemname-H's feature model, the data plane remains responsible solely for deterministic, streaming computation, while the control plane performs expressive ML analysis.
This implementation demonstrates that \systemname's architectural principles apply not only to statistical features, but also to distribution-based feature representations, without sacrificing Terabit-scale visibility.

%% file: sections/appendix-eval-meth.tex
\subsection{Experimental Setup, Datasets, and Metrics}
\label{appendix:eval-meth}

This appendix provides detailed information on the experimental setup and evaluation methodology used in Section~\ref{sec:evaluation}. 

\noindent
\textbf{Testbed.}
To evaluate \systemname's detection accuracy and runtime performance, we built a testbed composed of a Wedge 100BF-32X programmable switch equipped with an Intel Tofino 2 ASIC, and two commodity servers connected via 100 Gbps links.
Each server is equipped with dual-socket Intel Xeon Gold 6226R CPUs, 96 GB of DRAM, and Intel E810 NICs.
During evaluation, one server replays packet traces toward the switch, which executes packet processing and feature computation at line rate.
Feature records are then transmitted to a second server that runs the ML-based detection pipeline.

When measuring detection performance (\cref{subsec:eval-detection-kitsune}-~\cref{subsec:eval-detection-planter}), we replay traces at their original rates to preserve traffic characteristics.
When evaluating runtime performance (\cref{appendix:runtime-perf}), we instead use a DPDK-based packet generator to stress the pipeline at full line rate, decoupling performance evaluation from trace limitations.
For runtime measurements, a second switch supplies 100 Gbps of traffic to the device under test (DUT), configured with a snake topology. Traffic traverses the DUT repeatedly through successive port connections, exercising packet processing and feature computation at an aggregate rate of 12.8 Tbps.




\noindent
\textbf{Datasets.}
We evaluate \systemname using attack traces from three widely used datasets: Kitsune~\cite{mirsky2018kitsune}, CIC-IDS-2018~\cite{sharafaldin2018toward}, and HyperVision~\cite{fu2023hypervision}.
Together, these datasets contain 84 attacks spanning diverse behaviors, traffic volumes, and encryption scenarios, making them well-suited for evaluating robustness under aggressive sampling.
Kitsune and CIC-IDS-2018 are part of a collection of datasets commonly used to assess intrusion detection system performance, while the HyperVision dataset consists of labeled attacks generated in the authors' private cloud testbed across four categories: traditional brute-force attacks, encrypted flooding traffic, encrypted malicious web traffic, and malware-generated encrypted traffic.

For all baselines, models are trained exclusively on benign traffic drawn from the same dataset as the evaluated attack.
When evaluating the Kitsune and CIC-IDS-2018 datasets, we use the initial one million packets of each trace---comprised exclusively of benign traffic---for training.
The remaining portion of each trace contains malicious traffic associated with a specific attack and is used for evaluation.
For \systemname-H and the HyperVision baseline, we use the original HyperVision evaluation datasets and follow the methodology described in~\cite{fu2023hypervision}: models are trained on benign traffic to construct baseline interaction graphs, and sampling is applied only at inference time when evaluating detection performance.
By default, training is performed without sampling; sampling is applied only during testing unless stated otherwise.
This protocol measures the impact of introducing sampling after training.
~\cref{appendix:train-sampl} separately examines \systemname-K's sensitivity to sampling during training.

\noindent
\textbf{Sensitivity to aggregation parameters.}
For \systemname, the primary configuration parameter is the feature-record generation rate, controlled by epoch duration (\systemname-K) or flow-record emission frequency (\systemname-H).
Since this parameter directly determines the effective post-feature sampling rate, varying the sampling rate simultaneously captures the effect of these aggregation parameters.
We therefore report results as a function of sampling rate, which provides a unified view of the trade-off between traffic reduction and detection performance.



\noindent
\textbf{Evaluation metrics.}
\systemname-K produces an anomaly score based on reconstruction error, while \systemname-H outputs a clustering-based loss value associated with each flow.
In both cases, higher values indicate stronger deviation from learned benign behavior.
To compare detection effectiveness across systems and sampling rates without committing to a specific decision threshold, we use the Area Under the ROC Curve (AUC) as our detection performance metric.
AUC captures the trade-off between true-positive and false-positive rates across all possible thresholds, making it particularly appropriate for evaluating detectors under changing visibility conditions caused by sampling.
We additionally report F1-scores in selected analyses to confirm that the conclusions drawn from AUC remain unchanged.

%% file: sections/appendix-kitsune.tex
\subsection{Detection Performance against Kitsune}
\label{appendix:kitsune}

\begin{figure*}[t]
	\includegraphics[width=1\textwidth]{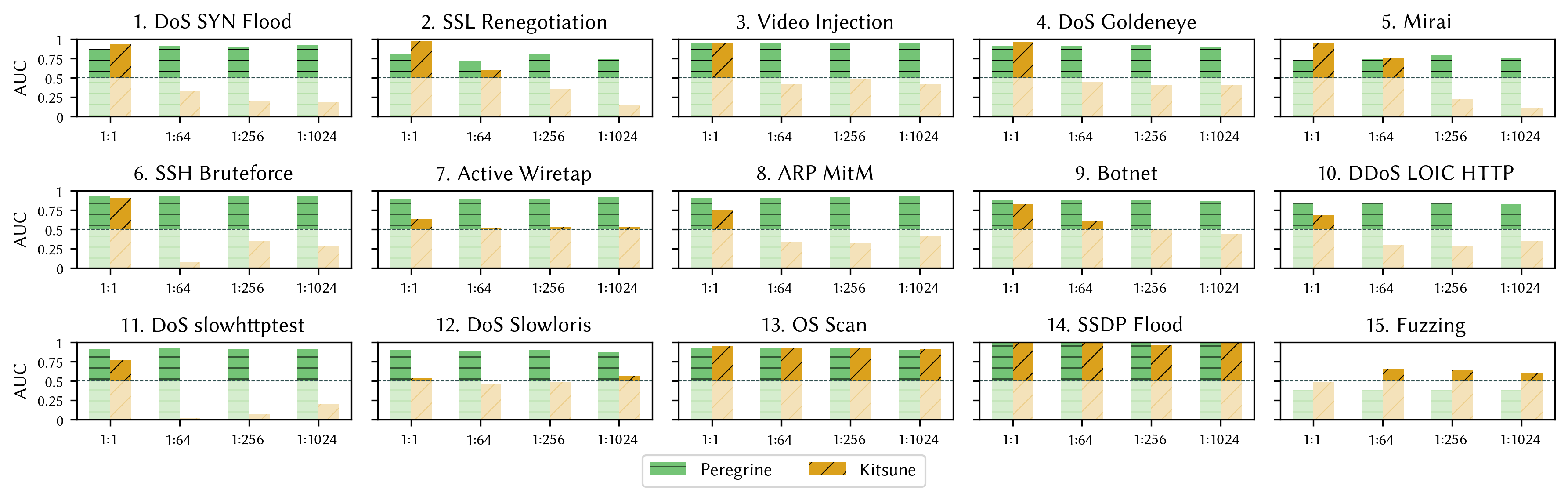}
	\caption{AUC across sampling rates (on x-axis): Kitsune and \systemname-K. At sampling rates representative of Terabit networks, \systemname-K is consistently more effective. At lower sampling rates, Kitsune remains effective for only a minority of attacks (3/15), whereas \systemname-K preserves high effectiveness for most (14/15).}
	\label{fig:kitnet-detection-auc}
\end{figure*}

This section compares \systemname-K's and Kitsune~\cite{mirsky2018kitsune}, contrasting feature computation in the data plane with feature computation in a middlebox server, using the original evaluation traces from~\cite{mirsky2018kitsune} and~\cite{cic-datasets} (15 attacks in total).
Figure~\ref{fig:kitnet-detection-auc} presents the results.
We evaluated each attack trace at various sampling rates to consider the downsampling of packet processing from a Tbps switch to a Gbps middlebox.

Both systems use the same ML backend, but differ in feature computation and sampling placement.
Kitsune samples packets before feature computation; \systemname-K computes features over all packets and applies sampling only when emitting feature records to the ML backend.

While the overall detection performance varies between attacks, consistently with other works~\cite{mirsky2018kitsune,whisperFu2021}, \systemname-K's is systematically more robust.
Kitsune performs well \textit{without} sampling but becomes ineffective (AUC $<$ 0.5) with sampling for 12/15 attacks, reflecting the visibility loss introduced by pre-feature packet sampling.
By contrast, \systemname-K retains strong performance (AUC $>$ 0.8) for 12/15 attacks and remains effective (AUC $>$ 0.5) for 14/15 attacks even at the lowest post-feature sampling rates.

We evaluate down to 1:1024 due to trace-length limitations.
The stability of \systemname-K's AUC from 1:1 down to 1:1024 indicates that sampling after feature computation preserves the behavioral signals needed by the ML model under the aggressive reduction factors required at Terabit scale.



Delving a bit into the details, we now divide the analysis of these results into three groups.

\noindent
\textbf{Attacks 1 to 5.} Without sampling (1:1), the performance of the Kitsune baseline for these attacks is slightly better than \systemname-K without sampling.
With sampling, however, Kitsune's performance falls abruptly.
\systemname-K, on the other hand, obtains a very strong performance across most attacks.
Crucially, its results \textit{are unaffected by sampling}.
This gap directly reflects the loss of per-flow statistics and temporal structure when packets are dropped before FC; when FC runs on all packets, the ML backend continues to observe stable summaries even if fewer records are exported.

\noindent
\textbf{Attacks 6 to 12.} Somewhat surprisingly, for these seven attacks \systemname-K outperforms Kitsune at every evaluated sampling rate, including the unsampled (1:1) setting.
Since both systems use the same ML backend, this difference cannot be attributed to changes in the detector itself. Instead, it likely reflects secondary effects introduced by the \systemname-K realization, such as data plane approximations, feature aggregation, or epoch-based summarization, which may smooth the input feature space and improve generalization~\cite{tibshirani1996regression,cortes2012l2,guo2018survey}.
We do not claim these effects as a primary contribution.
Rather, they may represent an implementation artifact that occasionally benefits detection.
Appendix~\ref{appendix:approx} analyzes these effects in more detail and shows that they are modest compared to the substantially larger gains observed under realistic sampling rates.
This supports our central conclusion that the dominant factor is architectural: preserving visibility by computing features before traffic reduction.

\noindent
\textbf{Attacks 13 to 15.} The results for these attacks are different from all of the above.
Both detectors perform well on OS Scan and SSDP Flood, likely due to the characteristics of the traces, where malicious traffic dominates benign traffic throughout the attack.
In contrast, neither system effectively detects the Fuzzing attack: its highly random inputs make it difficult for ML models to learn a consistent pattern.
This ``hard'' case helps contextualize our claim: post-feature sampling fixes the visibility bottleneck, but it does not eliminate fundamental limits of anomaly detection when the malicious behavior is diffuse and resembles high-entropy benign variability. 

\begin{figure*}[!t]
  \includegraphics[width=1\textwidth]{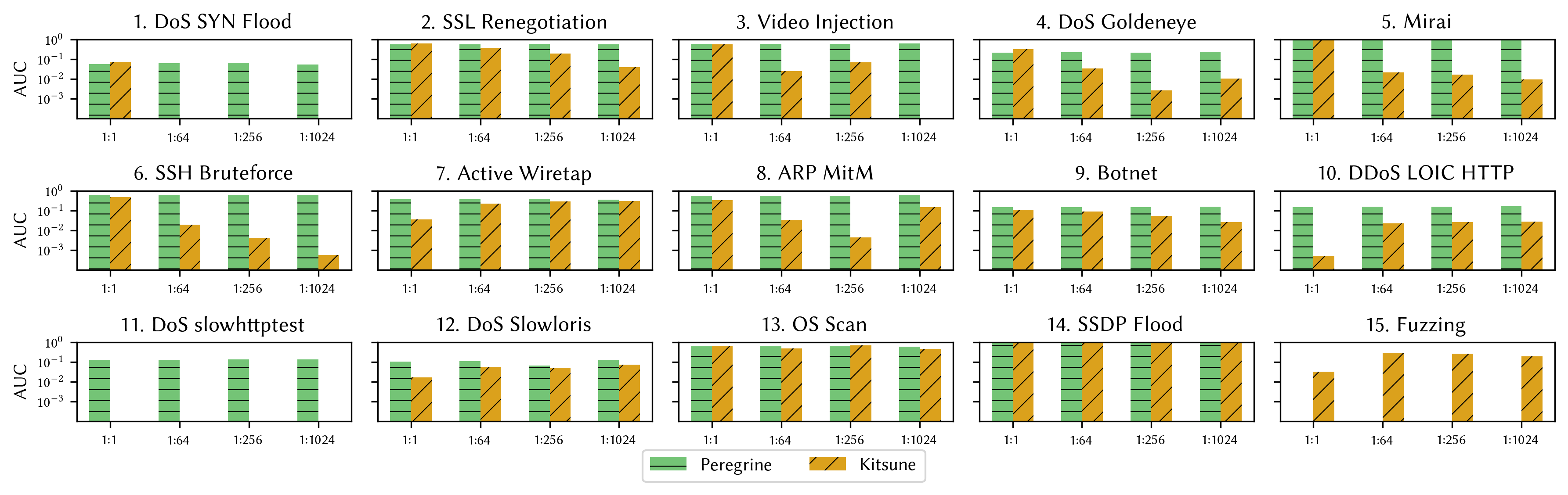}
  \caption{F1-score across sampling rates: Kitsune and \systemname-K. Threshold set to FPR=0.1.}
  \label{fig:kitnet-detection-f1-fpr-01}
\end{figure*}

\noindent
\textbf{Detection Performance against Kitsune: F1-score.}
The metric presented in the following section is the \textit{F1-score}, defined as the harmonic mean between \textit{precision} (proportion of instances identified as positives which are actually positives) and \textit{recall} (proportion of instances correctly identified as true positives).

\begin{figure*}[t]
  \includegraphics[width=1\textwidth]{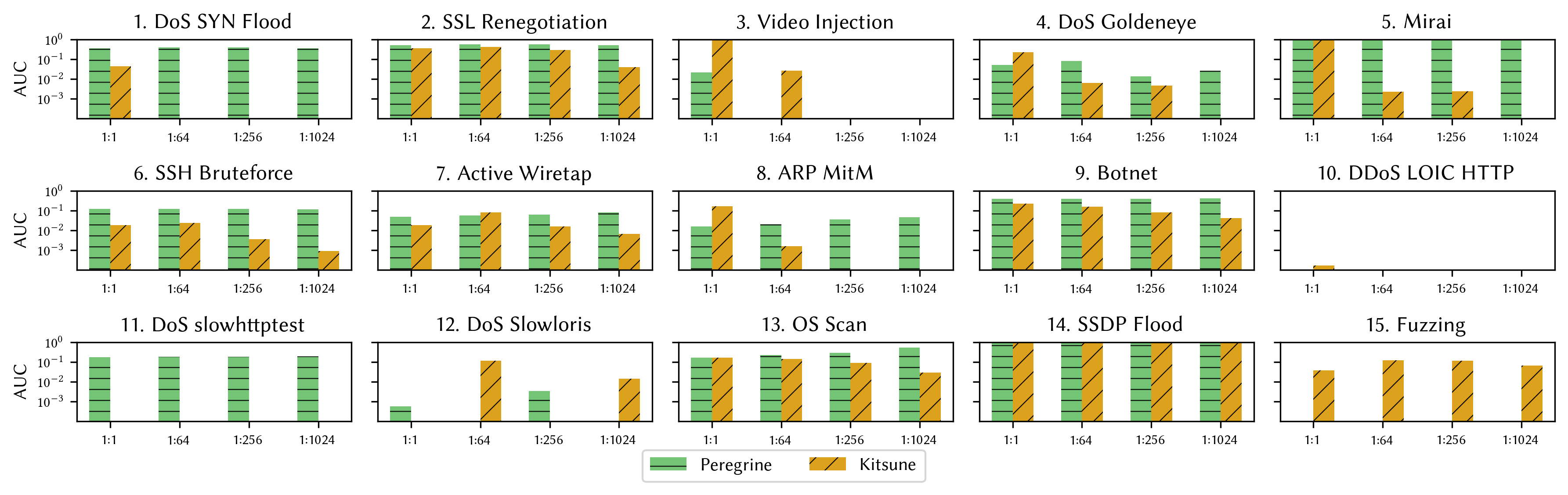}
  \caption{F1-score across sampling rates: Kitsune and \systemname-K. Threshold set to FPR=0.01.}
  \label{fig:kitnet-detection-f1-fpr-001}
\end{figure*}

A number of techniques can be used to select a threshold value for classification according to the needs of each operator (e.g., choosing a threshold value based on a maximum percentage of a given metric).
Figures~\ref{fig:kitnet-detection-f1-fpr-01} and~\ref{fig:kitnet-detection-f1-fpr-001} present results (logarithmic scale) for two such threshold values.
One which guarantees a False Positive Rate (FPR) of 0.1, a looser threshold that allows for more false positives to detect a greater number of attacks; and one with FPR = 0.01, a more conservative threshold that minimizes false positives.
The rationale for choosing these two values was to analyze the trade-off between maximizing attack detection and minimizing false alarms.

On the first case, in Figure~\ref{fig:kitnet-detection-f1-fpr-01}, as the threshold is set to a comparatively lower value, the range of packets identified by the model as outliers---either True Positives or False Positives---is higher.
While the results are generally high for both systems when the sampling is 1:1, \systemname-K achieves a higher F1-score than Kitsune on the remaining sampling rates, as expected.

In Figure~\ref{fig:kitnet-detection-f1-fpr-001}, when the threshold is set to FPR=0.01, the difference between the two systems becomes more noticeable.
While Kitsune often exhibits a clear decline in its results as the sampling rate increases (e.g., in the Mirai and SSH Bruteforce attacks), \systemname-K consistently maintains its detection performance across sampling rates in most attacks.
This latter property is in fact observed on both threshold values, highlighting \systemname-K's much stronger stability in terms of detection performance across different sampling rates.


%% file: sections/appendix-hv.tex
\subsection{Detection Performance against HyperVision}
\label{appendix:hv}






This section compares the detection performance of \systemname-H against HyperVision~\cite{fu2023hypervision}, using the original evaluation traces from~\cite{fu2023hypervision} (69 attacks in total).
This comparison is particularly instructive, as HyperVision represents the state of the art in high-performance, graph-based ML detection for multi-Gbps networks.
Figure~\ref{fig:hv-detection-auc-all} presents results for all attacks.


As expected, HyperVision generally outperforms \systemname-H in the unsampled (1:1) setting.
Even so, \systemname-H achieves equal or higher AUC scores for 22 of the 69 attacks, despite relying on data plane approximations and compact feature summaries.
As in~\cref{subsec:eval-detection-kitsune}, these occasional gains at 1:1 may reflect secondary smoothing or aggregation effects introduced by the \systemname-H realization.
However, our main conclusions do not depend on this observation and instead derive from the substantially larger differences observed under realistic sampling rates.

With sampling, the behavior of the two systems diverges sharply.
HyperVision's performance degrades rapidly as sampling rates decrease, reflecting the loss of structural and temporal information caused by pre-feature packet sampling.
In contrast, \systemname-H maintains effective detection performance across a substantial fraction of attacks even under aggressive post-feature sampling.

To make this contrast more concrete, we focus on the two lowest sampling rates evaluated (1:2048 and 1:4096), which correspond to realistic operating points when reconciling Terabit-scale data plane throughput with software-based ML inference.
At a sampling rate of 1:2048, HyperVision is ineffective (AUC $<$ 0.5) for 61 out of the 69 attacks.
Reducing the sampling rate further to 1:4096 nearly nullifies its effectiveness entirely, with only 2 out of 69 attacks achieving an AUC $>$ 0.5.
This collapse mirrors the behavior observed for Kitsune in \cref{subsec:eval-detection-kitsune} and highlights that even highly optimized detectors fundamentally rely on sufficient pre-feature visibility.

By contrast, \systemname-H remains effective (AUC $>$ 0.5) for 41 out of the 69 attacks across these same low sampling rates, demonstrating substantially greater robustness to aggressive downsampling.
Moreover, \systemname-H achieves strong detection quality (AUC $>$ 0.7) for 24 attacks, whereas HyperVision fails to reach an AUC above 0.7 for \textit{any} attack at the lowest evaluated sampling rate of 1:4096.

These results do not contradict HyperVision's strength in multi-Gbps settings; rather, they clarify its architectural limits.
HyperVision performs well with full packet visibility, but its detection performance degrades substantially under the evaluated pre-feature sampling regimes.
In contrast, \systemname-H is able to preserve detection effectiveness comparable to the unsampled scenario for a large fraction of attacks, precisely because sampling occurs after feature computation.

Overall, this comparison reinforces the paper's central claim: at Terabit scale, detection failure is driven not by insufficiently expressive ML models, but by the architectural decision to sample before feature-extraction.
State-of-the-art graph-based detectors collapse when deprived of packet-level visibility, whereas post-feature sampling preserves the behavioral signals required for effective ML-based detection.

%% file: sections/appendix-planter.tex

\subsection{Detection Performance against Planter}
\label{appendix:planter}

This section compares \systemname-K against Planter~\cite{zheng2024planter}, a state-of-the-art framework for deploying ML models directly in programmable switch data planes, using the evaluation traces from~\cite{mirsky2018kitsune,cic-datasets,fu2023hypervision} (84 attacks in total).
This comparison complements Sections~\ref{subsec:eval-detection-kitsune} and~\ref{subsec:eval-detection-hv}, and Appendices~\ref{appendix:kitsune} and~\ref{appendix:hv}: while those sections isolate the impact of sampling placement in software-based ML detectors, this section evaluates the alternative design point of executing ML inference entirely in the data plane.
Figure~\ref{fig:planter-detection-auc-all} reports results for all 84 attacks.

Planter enables the deployment of different ML algorithms to the switch using automatically generated P4 code and operator-configurable hyperparameters.
We evaluate Planter's Isolation Forest (IF) and Autoencoder (AE).

The evaluated Planter configurations use five-tuple fields as inputs.
To ensure that detection performance reflects behavioral anomalies rather than the appearance of previously unseen connections, the dataset is constructed so that benign and malicious traffic exhibit comparable connection dynamics.
This prevents trivial ``new-flow'' detection effects and isolates the models' ability to detect malicious behavior.

Following the original Planter configuration, the Isolation Forest model is deployed with 9 trees and 128 instances.
To evaluate the autoencoder-based model we use a modified version of KitNET's autoencoder that matches Planter's configuration (single hidden layer, no input feature normalization).

The results for \systemname-K correspond to a realistic post-feature sampling rate of 1:1024.
The Planter models use the 5-tuple as input features, reflecting both the original Planter evaluation methodology and the practical constraints of in-switch ML deployment.
We evaluate both baselines without input sampling.

The results show \systemname-K maintains effective detection for the vast majority of attacks (78/84).
Under the same attack set, \systemname-K outperforms Planter-IF on 62 of 84 attacks and Planter-AE on 69 of 84 attacks, despite Planter executing ML inference entirely in the data plane without sampling.
This trend is consistent across attack types and mirrors the limitations observed for other in-switch ML approaches: restricted feature expressiveness and shallow model capacity fundamentally constrain detection quality, even when full line-rate visibility is available.

Overall, this comparison highlights a fundamental trade-off.
Purely in-switch ML approaches eliminate sampling entirely but are intrinsically limited by the computational and memory constraints of programmable data planes.
\systemname occupies a different point in the design space: it preserves full traffic visibility and enables expressive ML inference by decoupling feature computation from detection, at the cost of higher decision latency.


%% file: sections/appendix-approx.tex
\subsection{The Impact of Approximated Computations}
\label{appendix:approx}

\begin{figure*}[!t]
  \includegraphics[width=1\textwidth]{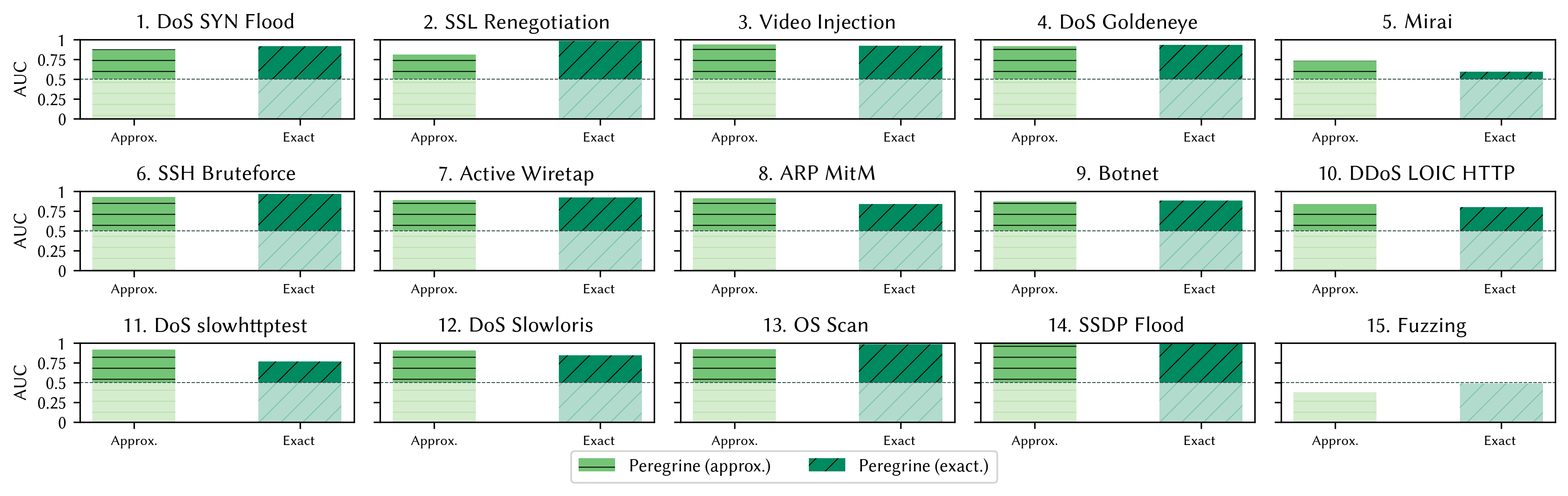}
  \caption{AUC score for a sampling rate of (1:1): \systemname-K. Detection performance comparison against a scenario where the input features are computed without approximations.}
 \label{fig:feature-error}
\end{figure*}

As described in Section~\ref{sec:implementation}, \systemname-K's feature computations incorporate a number of approximated computations.
To understand if these approximations affect the overall performance, we compared \systemname-K with a version with exact computations.
Figure~\ref{fig:feature-error} shows the obtained AUC scores, considering the Kitsune attack traces~\cite{mirsky2018kitsune}, but now comparing the performance with and without approximations, for a sampling rate of 1:1 (which allows us to focus only on the approximation factor).

Across most attacks, the results are nearly identical, indicating that the approximations introduced by \systemname-K do not meaningfully affect detection performance.
For a small number of attacks, the approximated version achieves slightly higher AUC scores.
While this may arise from minor changes in the feature values that occasionally benefit the downstream classifier, the observed improvements are modest.
More importantly, they are negligible compared to the substantially larger gains obtained under realistic sampling rates.
These results therefore suggest that the primary benefits of \systemname arise from preserving feature visibility under aggressive traffic reduction, rather than from any effects introduced by the approximations themselves.

%% file: sections/appendix-train-sampl.tex

\subsection{Detection Performance: Training with Sampling}
\label{appendix:train-sampl}

\begin{figure*}[!t]
  \includegraphics[width=1\textwidth]{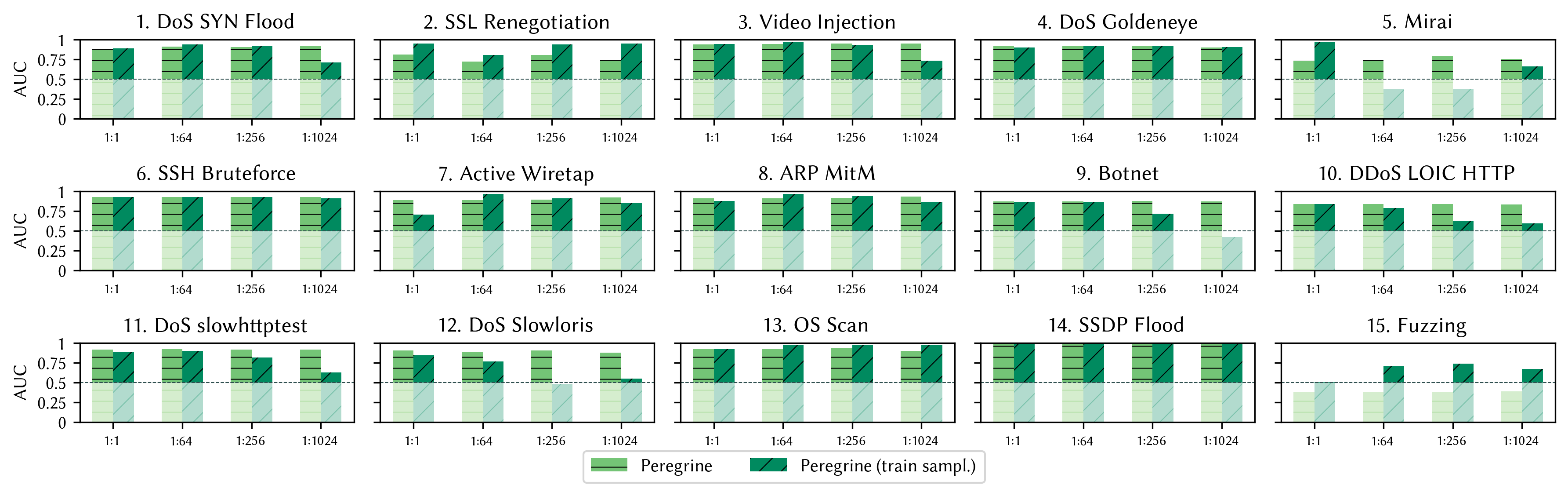}
  \caption{AUC across sampling rates: \systemname-K. Detection performance comparison against a scenario where sampling is performed during the training phase.}
  \label{fig:train-sampl}
\end{figure*}

As described in~\cref{appendix:eval-meth}, our detection performance evaluation was performed without sampling during the initial ML model training phase.
We now compare \systemname-K's detection performance with and without sampling during training.
The results are shown in Figure~\ref{fig:train-sampl}.

For most attacks, the performance of the ML model is not significantly affected by the use of sampling during training.
However, we observe that on some attacks the performance decreases when sampling is applied in the training phase.
As expected, by using sampling the training set is much smaller, leading to slightly less robust and effective models.

%% file: sections/appendix-runtime.tex
\subsection{Runtime Performance}
\label{appendix:runtime-perf}

\begin{figure}[t]
	\centering
	\includegraphics[width=.49\textwidth,keepaspectratio]{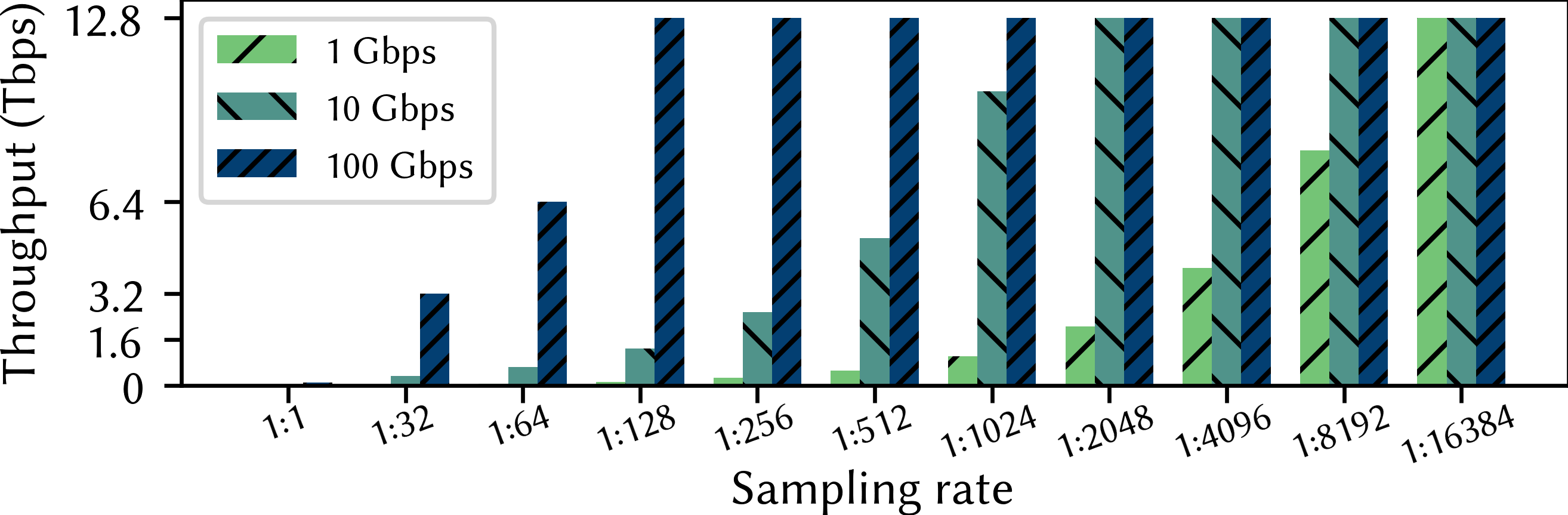}
	\caption{Throughput vs. sampling rate for different ML Detector modules.}
	\label{fig:perf_sampling_rate}
\end{figure}

This appendix complements the main evaluation by analyzing the runtime scalability of \systemname.
While Section~\ref{sec:evaluation} demonstrates that post-feature sampling preserves detection effectiveness at Terabit scale, this appendix quantifies the end-to-end throughput limits of the system and identifies remaining bottlenecks.

\noindent
\textbf{Runtime performance and sampling tradeo-ffs.}
As mentioned in ~\cref{subsec:rationale}, downsampling to the ML classifier is necessary, as server-side packet processing is several orders of magnitude slower than the network data plane.
Here we focus exclusively on throughput constraints imposed by server-side ML components and are agnostic to where sampling is applied; the impact of sampling placement on detection performance was analyzed separately in sections~\cref{subsec:eval-detection-kitsune}--~\cref{subsec:eval-detection-planter}.

The optimal sampling rate must balance two competing goals: high detection performance and ensuring that the ML detection (MD) component running on the middlebox server does not become a bottleneck.
ML detectors with poor runtime performance (i.e., low throughput) require low sampling rates to avoid limiting system throughput.
Conversely, higher-performance ML detectors allow higher sampling rates, potentially improving detection effectiveness.
Selecting an appropriate sampling rate therefore depends on the runtime capabilities of the ML component deployed in the control plane.

Figure~\ref{fig:perf_sampling_rate} illustrates how the expected throughput varies with the sampling rate for three ML detection modules with different runtime performances.
For an ML component capable of processing 1 Gbps of input traffic, \systemname requires a sampling rate of 1:16384 to avoid bottlenecking the server.
A detector sustaining 10 Gbps enables a higher sampling rate of 1:2048, while state-of-the-art ML systems capable of processing traffic close to 100 Gbps can support even higher sampling rates.

\begin{figure}[t]
	\centering
	\includegraphics[width=.49\textwidth,keepaspectratio]{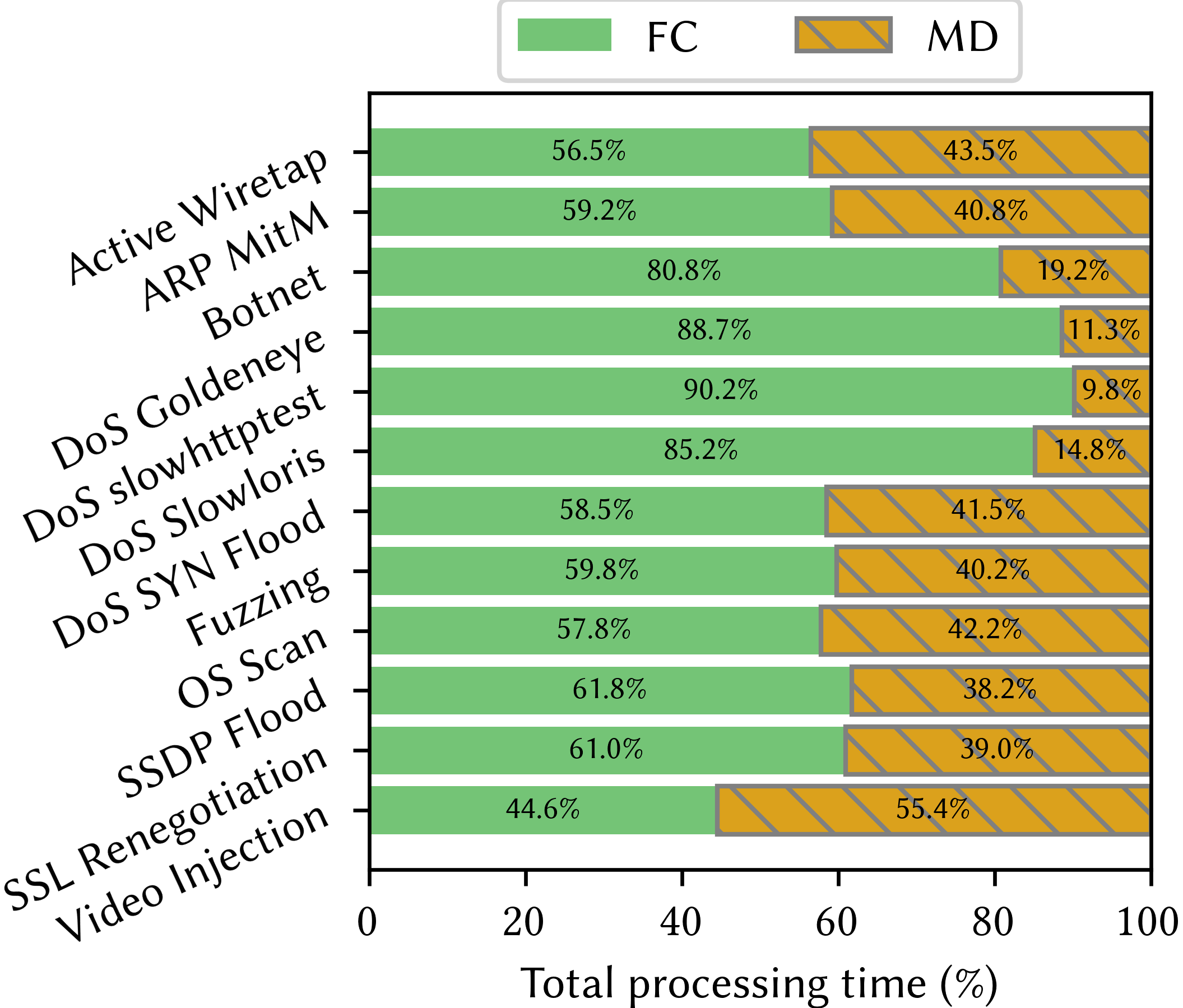}
	\caption{Relative weight of Feature Computation in the malicious traffic detection pipeline of \systemname-K.}
	\label{fig:fc_vs_ad}
\end{figure}

It is important to emphasize that Figure~\ref{fig:perf_sampling_rate} does not imply that \systemname processes traffic at the throughput shown for the MD component (e.g., 120~Gbps for \systemname-H).
These values correspond solely to the throughput of the server-side ML detector after traffic reduction.
Packet processing and feature computation are performed in the switch data plane at the full line rate of the target hardware (12.8 Tbps), and every packet contributes to the maintained feature state.
The lower MD throughput merely determines the amount of post-feature traffic reduction required to bridge the gap between Tbps packet processing and server-side ML inference.

\noindent
\textbf{Feature Computation: Relative Weight.}
As observed in Figure~\ref{fig:fc_vs_ad}, which showcases the percentage split between the FC and MD in terms of total processing time for multiple attacks, the FC component processing is, in most cases, heavier than MD.
Although it varies with each attack, it is commonly over 50\%, justifying the performance improvement in Figure~\ref{fig:pipeline_perf}.
This result is consistent with prior observations that feature-extraction can constitute a significant bottleneck in ML-based traffic analysis~\cite{bai2020fastfe}.

%% file: sections/appendix-resource-usage.tex
\subsection{Resource Usage}
\label{appendix:resource-usage}

\begin{table*}[]
\centering
\begin{tabular}{lcc|c|c|c|c|c|c|}
\cline{4-9}
\multicolumn{1}{c}{\textbf{}} & \textbf{} & \textbf{} & \textbf{Stages} & \textbf{Meter ALU} & \textbf{Hash Dist. Units} & \textbf{VLIW} & \textbf{SRAM} & \textbf{TCAM} \\ \hline
\multicolumn{1}{|l|}{\multirow{3}{*}{\textbf{\systemname-K}}} & \multicolumn{1}{c|}{\multirow{2}{*}{\textbf{TNA}}} & \textbf{Pipe 0} & 100\% & 72.9\% & 43.1\% & 26.6\% & 37.2\% & 6.9\% \\ \cline{3-9}
\multicolumn{1}{|l|}{} & \multicolumn{1}{c|}{} & \textbf{Pipe 1} & 91.7\% & 6.9\% & 0\% & 56.0\% & 6.4\% & 9.7\% \\ \cline{2-9}
\multicolumn{1}{|l|}{} & \multicolumn{1}{c|}{\textbf{TNA2}} &  & 95\% & 72.4\% & 26.3\% & 53.3\% & 26.9\% & 5.7\% \\ \hline
\multicolumn{1}{|l|}{\textbf{\systemname-H}} & \multicolumn{1}{c|}{\textbf{TNA2}} &  & 90\% & 86.3\% & 37.5\% & 13.9\% & 46.2\% & 54.2\% \\ \hline
\end{tabular}
\caption{\systemname: data plane resource usage (TNA and T2NA).}
\label{tab:peregrine-resource-usage}
\end{table*}

This appendix analyzes the data plane resource footprint of \systemname’s feature computation modules.
These results support the feasibility of computing expressive ML features at line rate on commodity programmable switches.

\systemname-K and \systemname-H's data plane resource usage on the Tofino Native Architecture (TNA) and Tofino 2 Native Architecture (T2NA) is summarized in Table~\ref{tab:peregrine-resource-usage}.
\systemname-K has both Tofino 1 (TNA) and Tofino 2 (T2NA) implementations, while \systemname-H has a single Tofino 2 implementation.

The TNA implementation is shown with the overall processing resources split between the two required switch pipelines, while the T2NA implementation is able to perform all feature computations in a single pipeline, thus avoiding packet recirculation.
This difference reflects the increased pipeline depth and resource availability of Tofino 2 compared to Tofino 1.

The main bottleneck of \systemname-K's TNA implementation is related to complexity of the feature computations required by \systemname-K.
As the feature computations performed require a significant number of operations, often involving multiple stateful memory accesses, the computation must be distributed across multiple processing stages and pipelines to fit within Tofino 1's architectural constraints, requiring packet recirculation.

Conversely, both \systemname-K and \systemname-H implementations for the Tofino 2 architecture perform all feature computations within a single pipeline, avoiding packet recirculation entirely.
This confirms that \systemname's feature set is compatible with modern programmable switch architectures and can be computed at the full line rate of the target hardware (12.8~Tbps) without compromising packet processing performance.

%% file: sections/appendix-efficiency.tex
\subsection{Cost Efficiency}
\label{appendix:efficiency}

\begin{figure*}[t]
  \centering
  \hspace{-1em}
  \subfigure[Power required at each traffic rate]{\includegraphics[width=.49\linewidth]{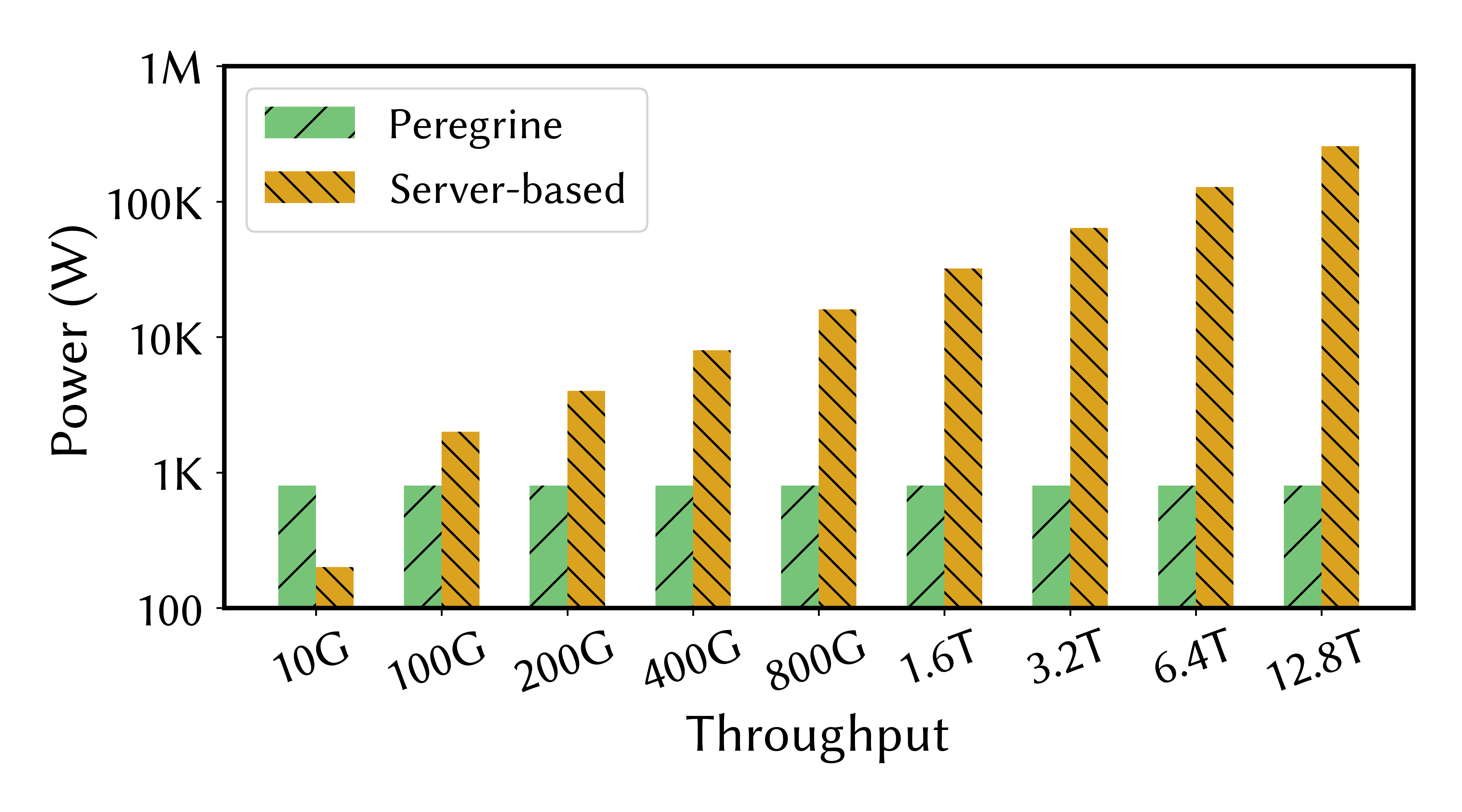}}
  \hspace{1em}
  \subfigure[Cost required at each traffic rate]{\includegraphics[width=.49\linewidth]{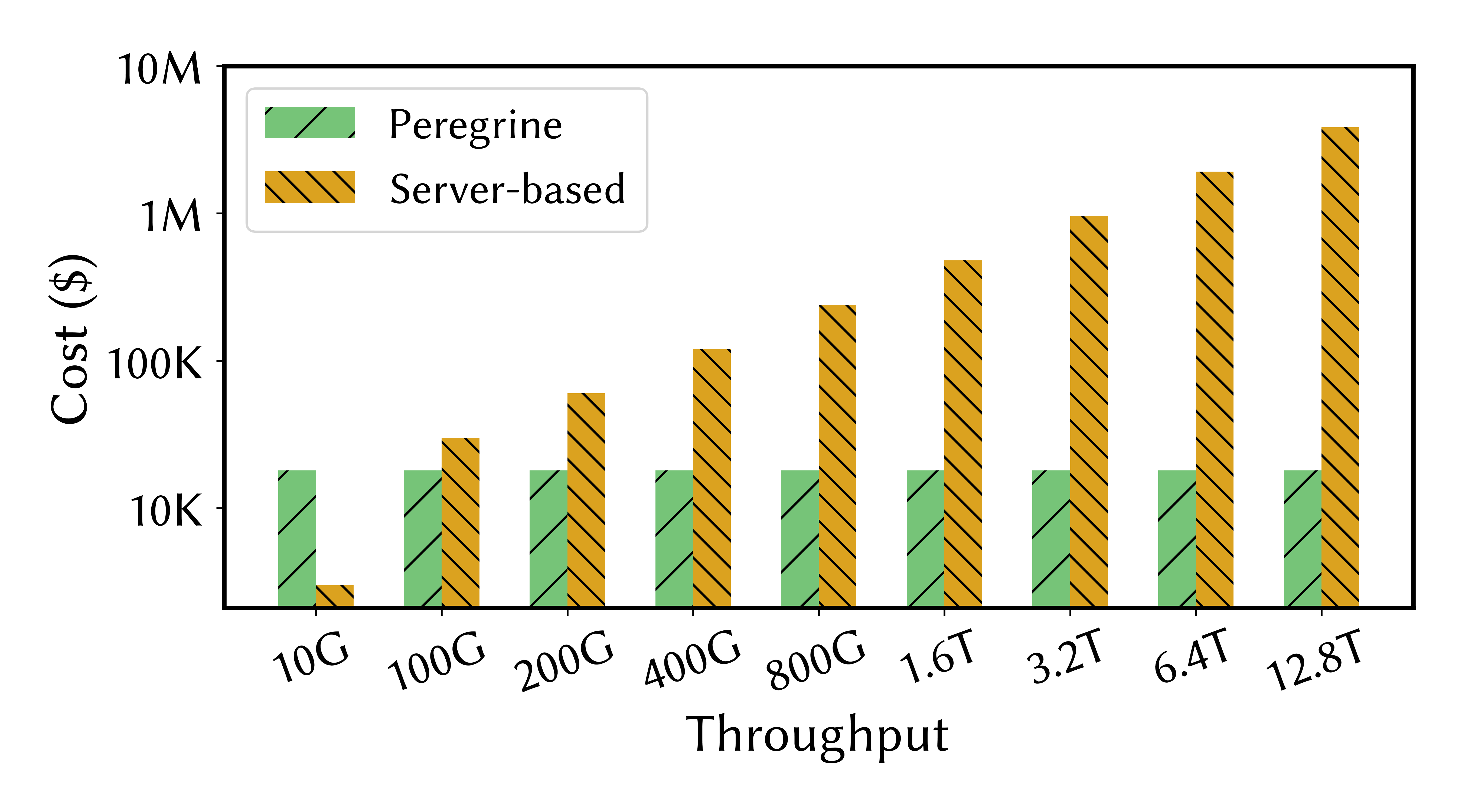}}
  \caption{Expected power consumption and cost of a server-based malicious traffic detector and \systemname~with increasing line rates.}
  \label{fig:cost_per_gbps}
\end{figure*}

We compare the cost efficiency of scaling ML-based malicious traffic detectors using commodity-server clusters against \systemname's HW-SW co-design.
Specifically, we estimate the monetary cost and power consumption of both approaches as traffic rates increase.

Our analysis is model-based but grounded in empirical data from both our evaluation and prior work~\cite{kim2020tea,liu2021jaqen,jacob2023does}.
For the server-based design point, we assume traffic is processed by a cluster of commodity servers that collectively sustain packet processing, feature computation, and ML inference.
The required number of servers is derived from the measured throughput of the evaluated detectors~\cite{mirsky2018kitsune,fu2023hypervision}.
Maintaining the same detection semantics as line rate increases therefore requires scaling the server cluster approximately proportionally to offered traffic.

For \systemname, packet processing and feature computation are performed in the programmable-switch data plane, while compact feature records are exported to a single inference server.
Switch power consumption is derived from published measurements~\cite{kim2020tea,liu2021jaqen,jacob2023does}.
To avoid overstating the benefits of our approach, we adopt conservative assumptions throughout and use the highest reported switch power-consumption figures when multiple values are available.
Consequently, the reported gains should be interpreted as conservative estimates of the potential cost and energy savings offered by \systemname.

\noindent
\textbf{Discussion.}
Figures~\ref{fig:cost_per_gbps}(a) and~\ref{fig:cost_per_gbps}(b) highlight a fundamental architectural distinction between the two approaches.
Higher-performance ML backends may allow server-based detectors to operate at higher sampling rates, but they do not eliminate the need to process packets and compute features in software.
As traffic rates increase, these costs scale approximately linearly with throughput, requiring increasingly large server deployments.
In contrast, \systemname performs per-packet processing and feature computation in the switch and exports only compact feature records to the ML backend.
As a result, the cost of packet processing and feature computation scales with switch capacity rather than server count, while the ML tier is dimensioned according to the desired feature-record sampling rate.
This reduces software-based per-packet processing, while backend demand and detection performance depend on the configured export rate.
Because the analysis intentionally adopts conservative assumptions, these results should be interpreted as estimates under the stated provisioning and sampling assumptions.

%% file: sections/appendix-rw.tex
\subsection{Related Work}
\label{appendix:rw}

Section~\ref{subsec:RW} and Table~\ref{tab:nids-comparison} summarize the systems most directly comparable to \systemname, focusing on scalability trade-offs.
This appendix discusses adjacent lines of work that are complementary to, or orthogonal to, that comparison, particularly programmable feature-extraction frameworks and telemetry systems that inform parts of our design.

\noindent
\textbf{In-network ML feature-extraction and streaming statistics}.
Stat4~\cite{gao2021stats} provides reusable P4 operators for computing streaming statistics such as means, variances, and percentiles.
Unlike \systemname, it exposes individual statistical primitives rather than an end-to-end ML-oriented feature-extraction pipeline with explicit post-feature sampling semantics.
P4DDLe~\cite{doriguzzi2023introducing}, Musumeci et al.~\cite{musumeci2022machine}, and FastFE~\cite{bai2020fastfe} similarly offload portions of feature-extraction to P4-programmable switches while performing classification in software.
FastFE completes feature computation on a server, while SuperFE~\cite{zhang2025superfe} batches packet metadata in a programmable switch and computes feature vectors on SmartNICs, demonstrating multi-100-Gbps operation on hardware.
These approaches accelerate feature-extraction; our focus is preserving detection effectiveness when downstream processing limits require a substantial traffic reduction.
In \systemname, feature computation precedes downsampling within the switch, so packets contribute to feature state even when no corresponding record is exported during their corresponding pipeline execution. The server backend therefore processes selected feature records rather than packet metadata from which it must construct those features.

\noindent
\textbf{In-network telemetry.}
Recent telemetry systems use query languages~\cite{narayana2017language} and cross-platform designs~\cite{gupta2018sonata,sonchack2018turboflow} to distribute monitoring functionality across servers and programmable switches.
Many systems further employ sketches~\cite{yu2013software,liu2016one,Sivaraman2017,yang2018elastic,tang2019mv,namkung2022sketchlib} to trade accuracy for memory efficiency.
Like \systemname, these systems use programmable switches to reduce the amount of information exported to software.
Their objective, however, is efficient telemetry and query execution rather than ML-based malicious traffic detection.
\systemname's key observation is that the placement of traffic reduction within the detection pipeline fundamentally affects detection effectiveness.
By computing ML-oriented behavioral features over all traffic before reduction occurs, \systemname preserves information that would otherwise be lost under pre-feature sampling.
Telemetry systems such as Sonata~\cite{gupta2018sonata} do not study this sampling-placement problem or its impact on downstream ML detectors.
While sketches and query-driven aggregation efficiently summarize traffic, they neither address detection-specific information loss caused by pre-feature sampling nor eliminate the need to reconcile Tbps packet rates with slower ML inference engines.

\noindent
\textbf{Host- and NIC-based acceleration.}
Systems such as Retina~\cite{Gerry2022Retina} and FAJITA~\cite{Ghasemirahni2024Fajita} improve packet-processing performance through optimized software pipelines and NIC-assisted execution, reaching tens to hundreds of millions of packets per second.
These approaches substantially reduce the gap between software analytics and network line rates, but still operate well below the packet-processing capabilities of modern programmable switches and must ultimately reconcile Tbps traffic rates with slower feature-extraction and ML inference.
Retina, for example, relies on early filtering to reduce processing load, discarding traffic before subsequent analysis.
In contrast, \systemname shifts feature computation into the switch data plane and applies traffic reduction only after feature-extraction, preserving visibility while reducing the load imposed on downstream ML engines.

\begin{figure*}[!t]
  \includegraphics[width=1\textwidth]{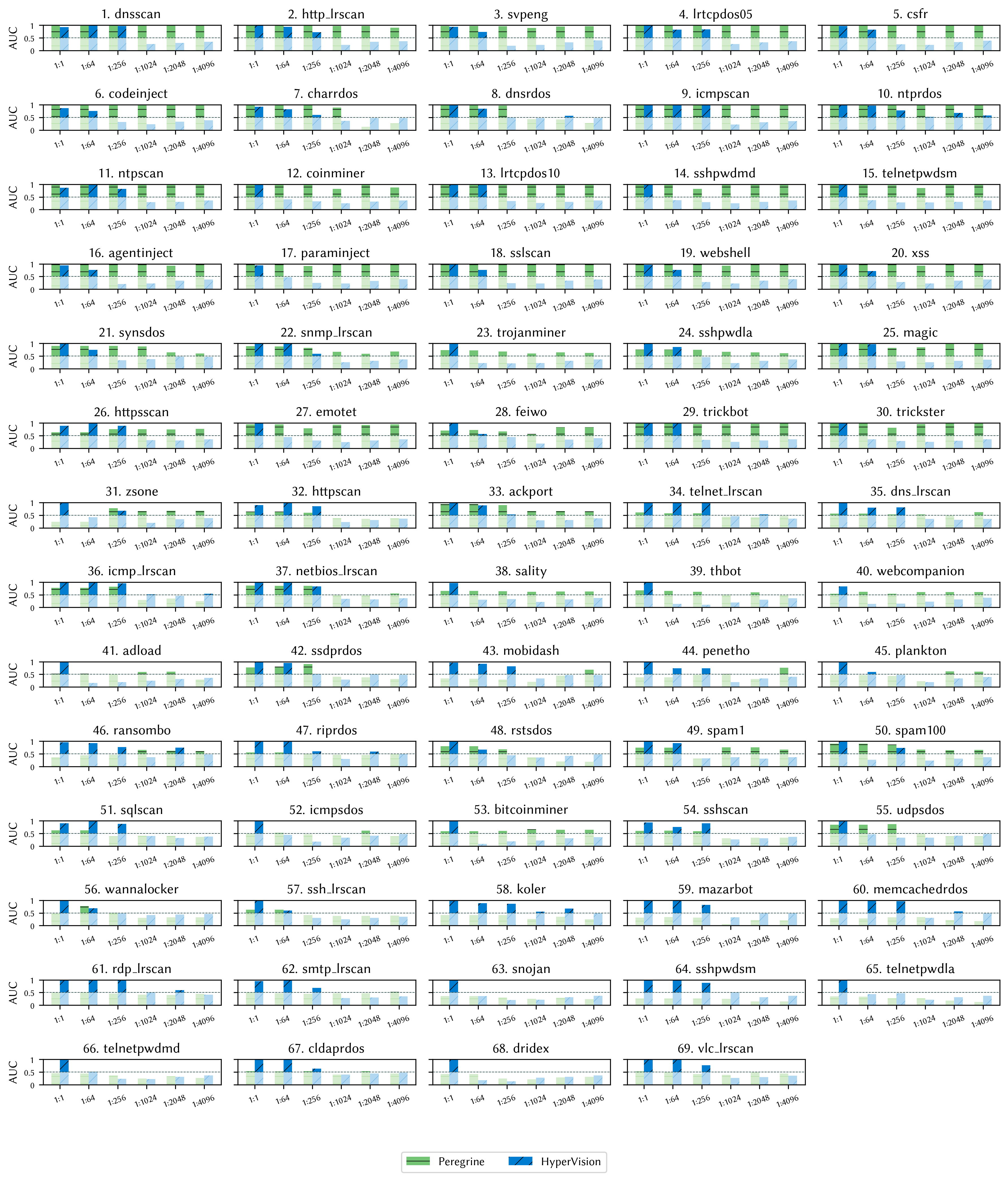}
  \caption{ AUC comparison between HyperVision and \systemname-H. At the lowest evaluated sampling rates, HyperVision falls below chance-level ranking (AUC $<$ 0.5) for the majority of attacks, whereas \systemname-H remains effective (AUC $>$ 0.5) for 41 of the 69 attacks. \systemname-H achieves strong detection quality (AUC $>$ 0.7) for 24 attacks, while HyperVision fails to achieve an AUC above 0.7 at the lowest evaluated sampling rate.}
  \label{fig:hv-detection-auc-all}
\end{figure*}

\begin{figure*}[!t]
  \includegraphics[width=1\textwidth]{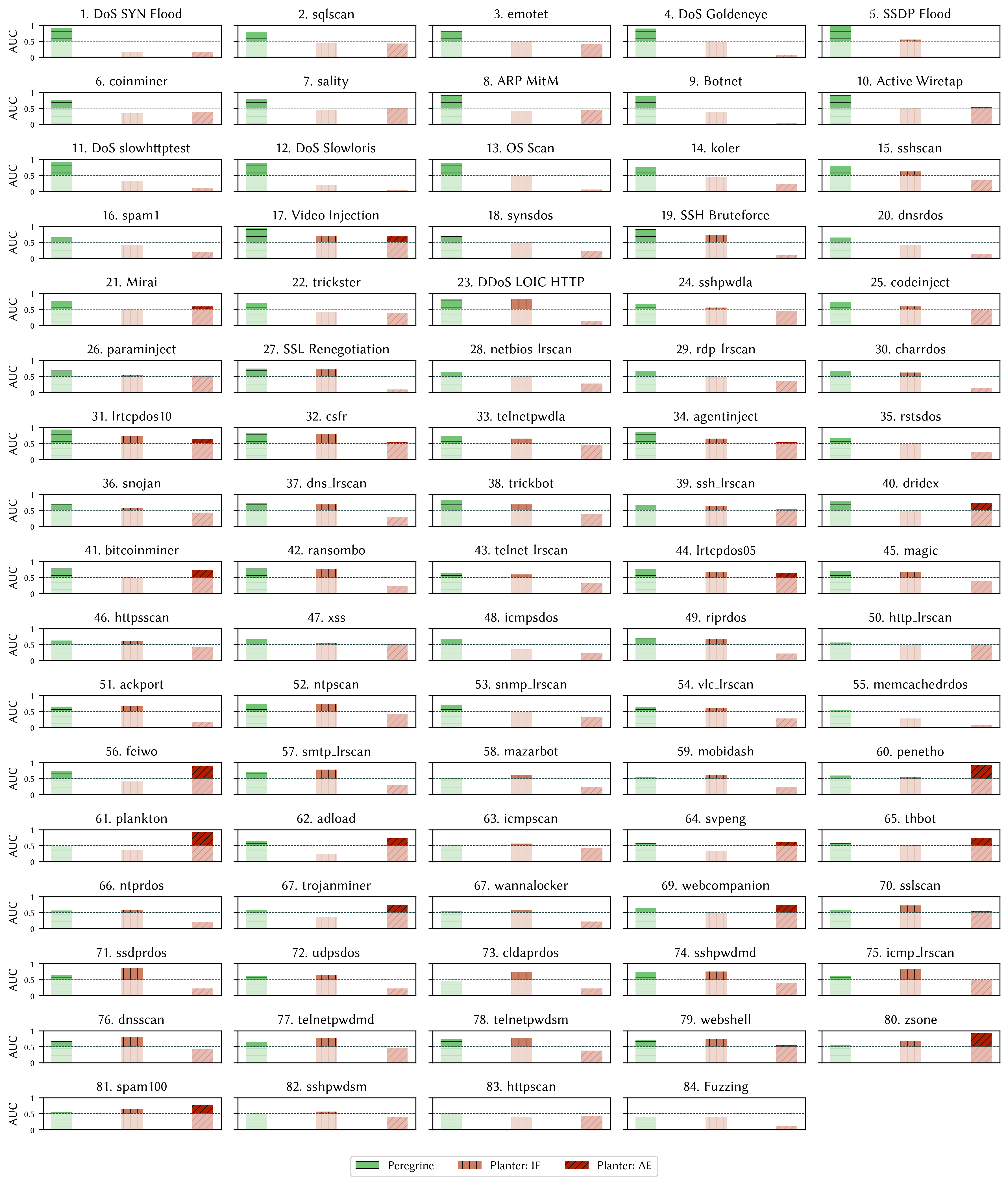}
  \caption{AUC comparison between the evaluated Planter-based baselines and \systemname-K. At realistic sampling rates (1:1024), \systemname-K remains effective for most attacks (78/84) and outperforms Planter-IF on 62/84 and Planter-AE on 69/84 attacks; Planter models are evaluated without sampling, as they execute ML inference in-network.}
  \label{fig:planter-detection-auc-all}
\end{figure*}